\documentclass[prb,preprint,preprintnumbers,amsmath,amssymb,floatfix]{revtex4}

\usepackage{graphicx}
\usepackage{dcolumn}

\begin{document}

\preprint{}

\title{A charge model as an effective model of one-dimensional Hubbard \\and extended Hubbard systems: its application to linear optical spectrum calculations in large systems based upon many-body Wannier functions}

\date{\today}

\author{S. Ohmura}
\affiliation{Nagoya Institute of Technology, Gokiso-cho, Syowa-ku, Nagoya 466-8555, Japan}
\author{A. Takahashi}
\affiliation{Nagoya Institute of Technology, Gokiso-cho, Syowa-ku, Nagoya 466-8555, Japan}
\author{K. Iwano}
\affiliation{Graduate University for Advanced Studies, Institute of Materials Structure Science,
High Energy Accelerator Research Organization (KEK), 1-1 Oho, Tsukuba 305-0801, Japan}
\author{T. Yamaguchi}
\affiliation{Institute of Materials Structure Science,
High Energy Accelerator Research Organization (KEK), 1-1 Oho, Tsukuba 305-0801, Japan}
\author{K. Shinjo}
\affiliation{Tokyo University of Science, 6-3-1 Niijuku, Katsushika-ku, Tokyo 125-8585, Japan}
\author{T. Tohyama}
\affiliation{Tokyo University of Science, 6-3-1 Niijuku, Katsushika-ku, Tokyo 125-8585, Japan}
\author{S. Sota}
\affiliation{Computational Materials Science Research Team, RIKEN Center for Computational Science (R-CCS), Kobe, Hyogo 650-0047, Japan}
\author{H. Okamoto}
\affiliation{Department of Advanced Materials Science, University of Tokyo, Chiba 277-8561, Japan}
\affiliation{AIST-UTokyo Advanced Operando-Measurement Technology Open Innovation Laboratory, 
National Institute of Advanced Industrial Science and Technology (AIST), Chiba 277-8568, Japan}

\begin{abstract}
We propose an effective model called the ``charge model", for the half-filled one-dimensional Hubbard and extended Hubbard models.
In this model, spin-charge separation, which has been justified from an infinite on-site repulsion ($U$) in the strict sense,
is compatible with charge fluctuations.
Our analyses based on the many-body Wannier functions succeeded in determining the optical conductivity spectra in large systems.
The obtained spectra reproduce the spectra for the original models well even in the intermediate $U$ region of $U=5$--$10T$, 
with $T$ being the nearest-neighbor electron hopping energy. These results indicate that the spin-charge separation works fairly well
in this intermediate $U$ region against the usual expectation and that the charge model is an effective model that applies to
actual quasi-one-dimensional materials classified as strongly correlated electron systems.
\end{abstract}

\maketitle
\section{Introduction}

The separation of spin and charge degrees of freedom (spin-charge separation) is considered to be a basic concept underpinning various properties of one-dimensional (1D) Mott insulators. The spin-charge separation was first recognized in Tomonaga--Luttinger liquids. In the weakly interacting 1D electron systems, collective excitations of charge and spin were shown to form instead of quasiparticles, and they are decoupled at low energies.\cite{TLL1,TLL2,TLL3,TLL4,TLL5,TLL6,TLL7} A liquid exhibiting this universal behavior is called a Tomonaga--Luttinger liquid. A power-law singularity of the momentum distribution function at the Fermi wave number and a power-law decay of the correlation functions originate from collective nature of excitations, and they are characteristics of Tomonaga--Luttinger liquids that distinguish them from Fermi liquids.\cite{TLL1,TLL2,TLL3,TLL4,TLL5,TLL6,TLL7} 
In the strong interaction limit, these spin and charge degrees of freedom were shown to separate in the ground state for the 1D Hubbard model at any filling for $U/T \to \infty$, $U$ being the on-site Coulomb interaction energy and $T>0$ the magnitude of the transfer integral.\cite{SCSWF} The origin of the spin-charge separation in the strong coupling case is different from that in the weak coupling case. In spite of the fact, the spin-charge-separated ground state has been shown to have the characteristic features of a Tomonaga--Luttinger liquid.\cite{SCSWF,SCSPL1,SCSPL2}

The ground state of the 1D Hubbard model is a Mott insulator at half-filling.
In the Mott insulators, an empty site (a holon, H) and a doubly occupied site (a doublon, D) are mobile excitations that may carry a charge. The ground state has neither Hs nor Ds in the limit $U/T \to \infty$; H and/or D can be generated though by chemical doping or photoexcitation. However, because chemical doping of the 1D Mott insulator materials is difficult, photoinduced phenomena are important stages to investigate the properties of these charge carriers. If the spin-charge degrees of freedom are separated, these charge carriers move freely without disturbing the spin state. Holon and spinon branches with different energy scales have been found in the angle-resolved photoemission spectrum, and this provides direct evidence of spin-charge separation.\cite{ARPES} Furthermore, spin-charge separation is considered to be the origin of novel optical properties of the 1D Mott insulators such as gigantic optical nonlinearity\cite{NLOS1,NLOS2,NLOS3,NLOS4} and the photoinduced transitions to metallic states.\cite{PIPT1,PIPT2}

The spin-charge separation has also been shown to hold for the photoexcited state in the limit $U/T \to \infty$.\cite{SCSEX1,SCSEX2,SCSOC1,SCSOC2} The density-density correlation function for the original extended Hubbard model, 
which is related to optical conductivity via the conservation of current,
is reproduced well by the spin-charge-separated photoexcited states.\cite{SCSOC2, stephan} 
Charge fluctuations are completely suppressed and the number of Hs and that of Ds are fixed to zero (one) in the ground state (photoexcited states) in the limit of $U/T \to \infty$. In the single hole case, the dynamical properties of the 1D Hubbard model are also known to originate from the spin-charge separation even considering charge fluctuations.\cite{FinU1,FinU2,FinU3}
In the optically excited states after the irradiation of visible or near-infrared light, on the other hand, charge fluctuations are expected to play the main
role.  Furthermore, we think that the degree of charge fluctuations will be substantial in 1D Mott insulator materials with typical $U/T$ values of 5--10.

To consider this problem, we introduce an effective model for the 1D Hubbard and extended Hubbard models, where spin-charge separation holds but charge fluctuations are not suppressed. The effective model is hereon called the charge model. By comparing the results obtained in the charge model with those in the original models, we can distinguish spin-charge coupling effects from charge fluctuation effects. We have found that the optical conductivity in the original models is reproduced quantitatively in the charge model despite the charge fluctuations significantly contributing to the optical conductivity in this realistic parameter range. The spin-charge separation and charge fluctuations are compatible in the 1D Mott insulators.

Femtosecond transient absorption spectroscopy has been a powerful experimental tool to investigate the physical properties of strongly correlated systems. As there exists no reliable approximation that can describe photoexcited states in the strongly correlated electron system, numerically exact diagonalization on small clusters\cite{ED} and the density matrix renormalization group (DMRG)\cite{DMRG,DDMRG,DDMRG2,DDMRG3,DDMRG4,DDMRG5,DDMRG6,DDMRG7,DDMRG8} are reliable theoretical methods to investigate transient absorption spectroscopy. However, finite-size effects are considerable in the exact diagonalization calculations. For example, because a band in the absorption spectrum of a macroscopic system changes to a few separated peaks in a small cluster, it is difficult to compare the absorption spectrum obtained by the exact diagonalization method with experimental results even if we introduce broadening to each peak. For larger system sizes, the absorption spectrum is calculated by the DMRG method, where finite-size effects are not significant. However, the wave functions of the ground state and photoexcited states are not obtained, and therefore interpreting the numerical results is difficult in this instance. As the dimension of the Hilbert space of the charge model is much smaller than that of the original Hubbard and extended Hubbard models, the charge model is a very good effective model to calculate the optical conductivity of a larger system. Furthermore, we propose a method to calculate the absorption spectrum and optical conductivity for these larger systems by introducing many-body ``Wannier functions'' (MBWFs), which are generated from linear combinations of energy eigenstates that have non-negligible transition dipole moments from the ground state. We have found that the optical conductivity calculated 
by the DMRG method is reproduced well even in a sufficiently large system 
in which finite-size effects are negligible.

The present paper is organized as follows. The charge model is introduced in Sec.~\ref{sec:CM}. The optical conductivity spectra calculated by the charge model is compared with that by the original Hubbard and extended Hubbard models in small clusters in Sec.~\ref{ssec:OCS}. 
 In Sec.~\ref{ssec:MBWFs}, we introduce a method using MBWFs to calculate the optical conductivity for a much larger system, and the optical conductivity spectra calculated by these two models are compared in sufficiently large systems that can be effectively regarded as the
 thermodynamic limit. In Sec.~\ref{sec:SD}, we give a brief summary and a discussion.
 Throughout this paper, we set $\hbar=e=1$ and lattice constant$=1$.
 
\section{Charge Model} \label{sec:CM}
The 1D extended Hubbard Hamiltonian describing the interaction of $N$ electrons at $N$ sites coupled to a light field is given by
\begin{eqnarray} \label{eq:eHH}
H(t)&=&{\hat K}(t)+{\hat V} \nonumber \\
{\hat K}(t)&=&\sum_{n=1}^{N}{\hat K}_n(t) \nonumber \\
{\hat K}_n(t)&=& -T \sum_{\sigma} \{c_{n, \sigma}^{\dagger} c_{n+1, \sigma} \exp[iA(t)] + {\rm H.c.} \}  \\
{\hat V}&=& U \sum_{n=1}^N c_{n, \uparrow}^{\dagger} c_{n, \uparrow} c_{n, \downarrow}^{\dagger} c_{n, \downarrow}
+ V \sum_{n=1}^N \sum_{\sigma,\sigma^{\prime}} c_{n, \sigma}^{\dagger} c_{n, \sigma} c_{n+1, \sigma^{\prime}}^{\dagger} c_{n+1, \sigma^{\prime}}. \nonumber
\end{eqnarray}
The term ${\hat K}(t)$ describes the transfer of electrons, where $c_{n, \sigma}^{\dagger}$ ($c_{n, \sigma}$) creates (annihilates) an electron of spin $\sigma$ at site $n$, and $A(t)$ is the dimensionless vector potential at time $t$. The electron--field coupling has been introduced into the transfer integral as a Peierls phase. The term ${\hat V}$ describes the Coulomb interaction, where $V$ is the Coulomb interaction energy between neighboring sites. A periodic boundary condition is imposed in that $c_{N+1, \sigma}=c_{1, \sigma}$ holds.

We construct an effective model for the half-filled 1D Hubbard and extended Hubbard models in a subspace $S$ spanned by the following basis states,
\begin{eqnarray} \label{eq:bases}
&& |\{p_1,p_2,\cdots,p_M\},\{q_1,q_2,\cdots,q_M\}\rangle  =D_{p_1}^{\dagger}D_{p_2}^{\dagger} \cdots D_{p_M}^{\dagger} \nonumber \\
&\times& \sum_{\sigma_1,\sigma_2,\cdots,\sigma_{N-2M}}f^{(M)}(\sigma_1,\sigma_2,\cdots,\sigma_{N-2M})
c_{l_1, \sigma_1}^{\dagger}c_{l_2, \sigma_2}^{\dagger}\cdots c_{l_{N-2M}, \sigma_{N-2M}}^{\dagger}|0\rangle,
\end{eqnarray}
where $M$ is the number of H-D pairs, $p_1<p_2<\cdots<p_M$, $q_1<q_2<\cdots<q_M$, and $l_1<l_2<\cdots<l_{N-2M}$ show doubly occupied, empty, and singly occupied sites, respectively, $D_{p}^{\dagger}=c_{p, \uparrow}^{\dagger}c_{p, \downarrow}^{\dagger}$ creates D at site $p$, and $|0\rangle$ is the vacuum state.
The spin wave function $f^{(M)}$ is independent of the charge configuration $\{p_1,p_2,\cdots,p_M\}$ and $\{q_1,q_2,\cdots,q_M\}$, and all the basis states with the same $M$ have the same spin wave function. The spin and charge degrees of freedom are separated in all the states in $S$ because of this property.
The spin wave function $f^{(M)}$ is given by the ground state of the Heisenberg Hamiltonian with $N-2M$ sites. The ground state of the charge model is given by the spin wave function $f^{(0)}$ in the limit $T/(U-V) \to 0$. Therefore, the ground state of the original extended Hubbard Hamiltonian can be reproduced in the charge model in the strong-coupling limit, which justifies the choice of $f^{(M)}$. The ground state of the Heisenberg Hamiltonian satisfies the cyclic condition,
\begin{eqnarray} \label{eq:TS}
f^{(M)}(\sigma_{1},\sigma_{2},\cdots,\sigma_{N-2M})
=\exp(-i\theta_{M})f^{(M)}(\sigma_{2},\cdots,\sigma_{N-2M},\sigma_{1}),
\end{eqnarray}
where a constant $\theta_{M}$ is given by
\begin{eqnarray} \label{eq:thetaM}
\theta_{M}=\frac{\pi}{2}{\rm mod}(N-2M,4),
\end{eqnarray}
and ${\rm mod}(N-2M,4)$ is the remainder of $(N-2M)/4$.
The basis states are normalized and satisfy the condition:
\begin{eqnarray} \label{eq:normalization}
\langle\{p_1^{\prime},p_2^{\prime},\cdots,p_M^{\prime}\},\{q_1^{\prime},q_2^{\prime},\cdots,q_M^{\prime}\}
|\{p_1,p_2,\cdots,p_M\},\{q_1,q_2,\cdots,q_M\}\rangle \nonumber \\
=\delta_{p_1^{\prime},p_1}\delta_{p_2^{\prime},p_2}\cdots \delta_{p_M^{\prime},p_M}
\delta_{q_1^{\prime},q_1}\delta_{q_2^{\prime},q_2}\cdots \delta_{q_M^{\prime},q_M}.
\end{eqnarray}

We consider an effective Hamiltonian given by
\begin{eqnarray} \label{eq:CHP}
H^{\rm (C)}(t)= PH(t)P,
\end{eqnarray}
where $P$ is a projection operator onto the subspace $S$. The model described by the effective Hamiltonian is termed as a charge model.
Since $S$ is invariant under ${\hat V}$, the effective Hamiltonian can be written as
\begin{eqnarray} \label{eq:CMH}
H^{\rm (C)}(t) &=&{\hat K}^{\rm (C)}(t)+{\hat V}, \nonumber \\
{\hat K}^{\rm (C)}(t)&=&\sum_{n=1}^{N}{\hat K}^{\rm (C)}_n(t), \\
{\hat K}^{\rm (C)}_n(t)&=&P{\hat K}_n(t)P. \nonumber
\end{eqnarray}

\begin{figure}[thbp]\centering
  \includegraphics[width=120mm]{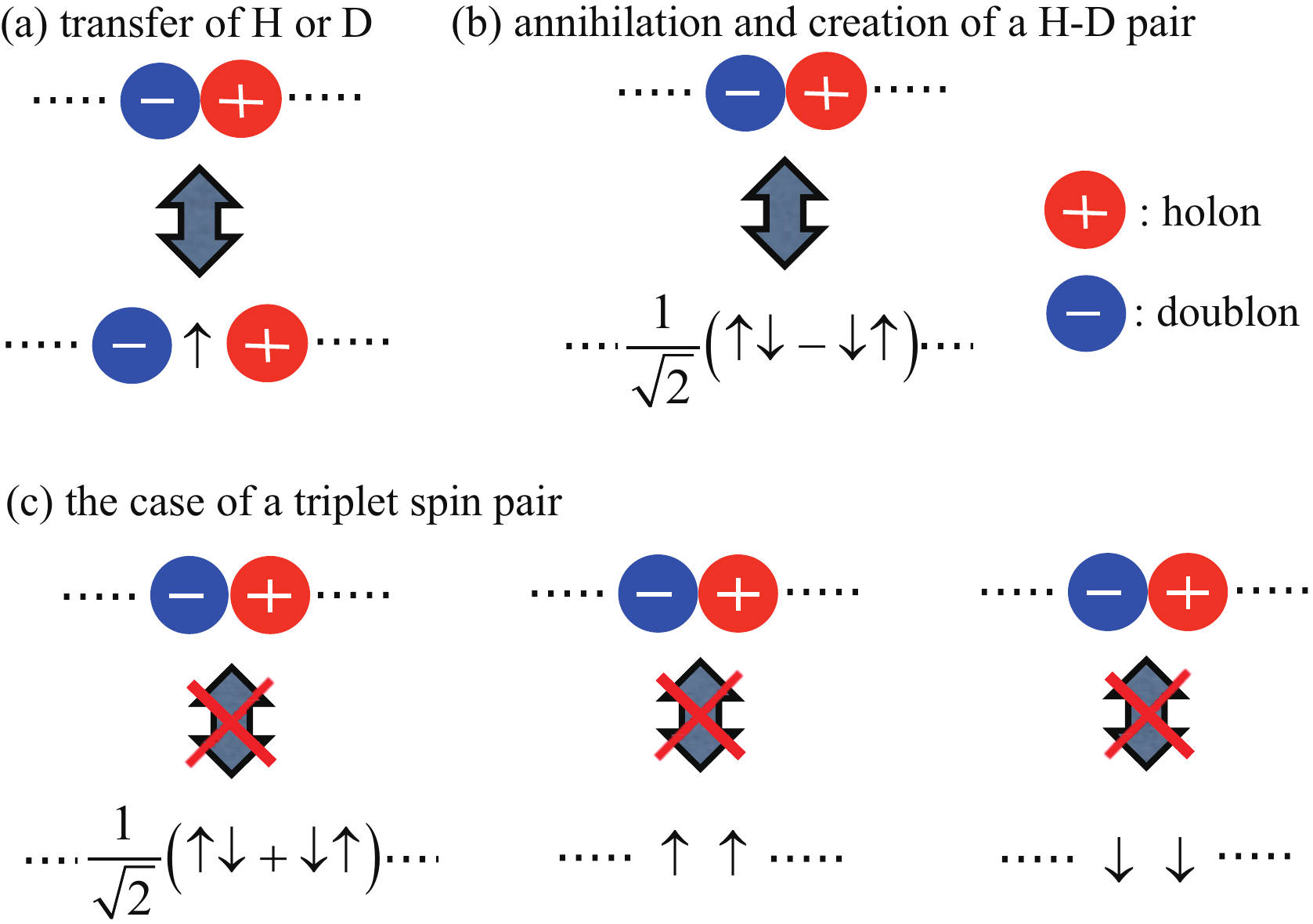}
  \caption{The possible change patterns of the electronic configuration at sites $n$ and $n+1$ by operating with ${\hat K}_n(t)$. (a) An example of the transfer of an H or D and (b) annihilation and creation of an H-D pair involved with a singlet spin pair are permitted. In contrast, (c) annihilation and creation of an H-D pair involved with a triplet spin pair are forbidden.}
  \label{fig:changingP}
\end{figure}

To derive ${\hat K}^{\rm (C)}_n(t)$, we show how the electronic configuration at sites $n$ and $n+1$ changes by operating with ${\hat K}_n(t)$ on states $|\cdots{\rm X}_n{\rm X}_{n+1}\cdots\rangle$ for which the electronic configuration at site $n$ is ${\rm X}_n$ and that at site $n+1$ is ${\rm X}_{n+1}$. Specifically, ${\rm X}_n=\sigma$ indicates that site $n$ is singly occupied with spin $\sigma$, and ${\rm X}_n={\rm D}$ (${\rm X}_n={\rm H}$) indicates that site $n$ is doubly occupied (empty).
The explicit expressions are given in Appendix~\ref{app1}.
The possible change patterns are as follows, and they are schematically shown in Fig.~\ref{fig:changingP}.

(i) Transfer of an H or D,
\begin{eqnarray} \label{eq:HDtransfer}
{\hat K}_n(t) |\cdots{\rm D}\sigma\cdots\rangle 
&=& Te^{-iA(t)}|\cdots\sigma{\rm D}\cdots\rangle,  \nonumber \\
{\hat K}_n(t) |\cdots\sigma{\rm D}\cdots\rangle 
&=& Te^{iA(t)}|\cdots{\rm D}\sigma\cdots\rangle,  \nonumber \\
{\hat K}_n(t) |\cdots{\rm H}\sigma\cdots\rangle 
&=& -Te^{iA(t)}|\cdots\sigma{\rm H}\cdots\rangle, \\
{\hat K}_n(t) |\cdots\sigma{\rm H}\cdots\rangle 
&=& -Te^{-iA(t)}|\cdots{\rm H}\sigma\cdots\rangle.  \nonumber
\end{eqnarray}
(ii) Annihilation of an H-D pair,
\begin{eqnarray} \label{eq:dHDp}
{\hat K}_n(t) |\cdots{\rm DH}\cdots\rangle 
&=& Te^{iA(t)}(|\cdots\uparrow\downarrow\cdots\rangle-|\cdots\downarrow\uparrow\cdots\rangle),  \nonumber \\
{\hat K}_n(t) |\cdots{\rm HD}\cdots\rangle
&=& Te^{-iA(t)}(|\cdots\uparrow\downarrow\cdots\rangle-|\cdots\downarrow\uparrow\cdots\rangle).
\end{eqnarray}
(iii) Creation of an H-D pair from a singlet spin pair,
\begin{eqnarray} \label{eq:cHDp}
{\hat K}_n(t) \frac{1}{\sqrt 2}(|\cdots\uparrow\downarrow\cdots\rangle-|\cdots\downarrow\uparrow\cdots\rangle)
= -{\sqrt 2}T(e^{iA(t)}|\cdots{\rm DH}\cdots\rangle+e^{-iA(t)}|\cdots{\rm HD}\cdots\rangle).
\end{eqnarray}
For a triplet pair, the following relation holds,
\begin{eqnarray} \label{eq:triplet}
{\hat K}_n(t) \frac{1}{\sqrt 2}(|\cdots\uparrow\downarrow\cdots\rangle+|\cdots\downarrow\uparrow\cdots\rangle)
&=& {\hat K}_n(t) |\cdots\uparrow\uparrow\cdots\rangle = {\hat K}_n(t)|\cdots\downarrow\downarrow\cdots\rangle=0,
\end{eqnarray}
In the case of transfer of an H or D, only the position of an H or D is changed but the spin wave function is not. The expressions of ${\hat K}^{\rm (C)}_n(t)$ in this case are explicitly given in Appendix~\ref{app1} as Eqs.~(\ref{eq:CMH1}--\ref{eq:CMH4},\ref{eq:CMHN1}--\ref{eq:CMHN4}). A phase factor appears when an H or D crosses the boundary. This is because the creation operators for the singly occupied sites are rearranged from left to right in increasing order of $l_k$.

In the case of the annihilation of an H-D pair, the spin wave function is changed.
We consider a state $|\{p_1,p_2,\cdots,p_{M}\},\{q_1,q_2,\cdots,q_{M}\}\rangle$ with $M$ H-D pairs, where a H and a D exist at sites $n$ and $n+1$, respectively, and $l_k<p_i=n$ and $q_j=n+1<l_{k+1}$ hold.
The H-D pair is converted to a singlet spin pair by operating with ${\hat K}_n(t)$. Therefore, the transferred state 
\begin{eqnarray} \label{eq:Knbase}
|\Phi_n\rangle={\hat K}_n(t) |\{p_1,p_2,\cdots,p_M\},\{q_1,q_2,\cdots,q_M\}\rangle
\end{eqnarray}
is given by
\begin{eqnarray} \label{eq:baseFHDPD}
|\Phi_n\rangle &=& -2Te^{iA(t)} D_{p_1}^{\dagger} \cdots D_{p_{i-1}}^{\dagger} D_{p_{i+1}}^{\dagger} \cdots D_{p_M}^{\dagger}
\sum_{\sigma_1,\sigma_2,\cdots,\sigma_{N-2M}}
f^{(M)}(\sigma_1,\sigma_2,\cdots,\sigma_{N-2M}) \nonumber \\
&\times& c_{l_1, \sigma_1}^{\dagger}\cdots c_{l_{k}, \sigma_{k}}^{\dagger}
(c_{n, \uparrow}^{\dagger}c_{n+1, \downarrow}^{\dagger}-c_{n, \downarrow}^{\dagger}c_{n+1, \uparrow}^{\dagger})
c_{l_{k+1}, \sigma_{k+1}}^{\dagger}\cdots c_{l_{N-2M}, \sigma_{N-2M}}^{\dagger}|0\rangle.
\end{eqnarray}
The spin wave function of $|\Phi_n\rangle$ is obtained by inserting a nearest-neighbor singlet pair of spins into the ground state $|\psi_0^{\rm (H)}(N-2M)\rangle$ of the 1D Heisenberg Hamiltonian with $N-2M$ sites between sites $l_k$ and $l_k+1$. The singlet spin pair (SSP) inserted state is denoted by $|\psi_0^{\rm (H)}(N-2M)+{\rm SSP}\rangle$. The spin wave function $|\psi_0^{\rm (H)}(N-2M)+{\rm SSP}\rangle$ is different from $|\psi_0^{\rm (H)}(N-2M+2)\rangle$, showing that spin-charge coupling is induced by the annihilation process of an H-D pair. The overlap between $|\psi_0^{\rm (H)}(N-2M+2)\rangle$ and $|\psi_0^{\rm (H)}(N-2M)+{\rm SSP}\rangle$ can be written as
\begin{eqnarray} \label{eq:OL}
\langle \psi_0^{\rm (H)}(N-2M)+{\rm SSP}|\psi_0^{\rm (H)}(N-2M+2)\rangle= c_{\rm S}(M) \exp[i(\theta_{M-1}-\theta_{M})(k-1)].
\end{eqnarray}
Note that the phases of these two states may be chosen independently, and that the overlap is multiplied by $\exp[i(\theta_{M-1}-\theta_{M})]$ if the location of the singlet spin pair is shifted by one site. We have chosen the phases so that the overlap is real and positive when the singlet spin pair is inserted at the first two sites in $|\psi_0^{\rm (H)}(N-2M)+{\rm SSP}\rangle$, and the overlap in this case is denoted by $c_{\rm S}(M)$. The value $c_{\rm S}^2(M)$ shows the weight of the singlet component of a spin pair at neighboring two sites in $|\psi_0^{\rm (H)}(N-2M+2)\rangle$. The system size dependence of the overlap was calculated, and it has been shown that $c_{\rm S}(M)$ are well fitted by the function $0.820+0.740(N-2M)^{-2}$.\cite{OLHG} We neglect the system size dependence, and adopt the value in the thermodynamic limit ($c_{\rm S}(M)=0.82$) for simplicity.

Using Eqs.~(\ref{eq:baseFHDPD}) and (\ref{eq:OL}), the only non-zero matrix element of ${\hat K}_n(t)$ within the subspace $S$ with column index $|\{p_1,p_2,\cdots,p_M\},\{q_1,q_2,\cdots,q_M\}\rangle$ is given by
\begin{eqnarray} \label{eq:baseFHDPD2}
\langle\{p_1,\cdots,p_{i-1},p_{i+1},\cdots,p_M\},\{q_1,\cdots,q_{j-1},q_{j+1},\cdots,q_M\}|{\hat K}_n(t)
|\{p_1,p_2,\cdots,p_M\},\{q_1,q_2,\cdots,q_M\}\rangle \nonumber \\
=-2Te^{iA(t)} c_{\rm S}(M) \exp[i(\theta_{M-1}-\theta_{M})(k-1)].\hspace{15pt}
\end{eqnarray}
When a D and a H exist at sites $n$ and $n+1$, respectively, we can obtain the matrix elements from the same procedure.
 The expressions of ${\hat K}^{\rm (C)}_n(t)$ in this case are explicitly given in Eqs.~(\ref{eq:CMH6},\ref{eq:CMH7},\ref{eq:CMHN6},\ref{eq:CMHN7}).

Creation of an H-D pair is the inverse process of annihilation of an H-D pair. Using this fact, ${\hat K}^{\rm (C)}_n(t)$ are obtained as explicitly given in Eqs.~(\ref{eq:CMH5},\ref{eq:CMHN5}). The absolute values of the matrix elements are reduced by the factor $c_{\rm S}(M)$ when the number of H-D pairs is changed. 

The constants $\theta_{M}$ and $c_{\rm S}(M)$ depend on the spin wave function, and the optical properties in the charge model depend on the spin wave function only through $\theta_{M}$ and $c_{\rm S}(M)$. Optical properties have been investigated in the limit $U/T \to \infty$, where the charge fluctuations were neglected and only the states with one H-D pair ($M=1$) were considered.\cite{stephan} They considered different spin wave functions and the contribution of the spin wave functions with various $\theta_{1}$ were considered there. The spin wave function with $\theta_{1}=({\pi}/{2}){\rm mod}(N-2,4)$ has been shown to have dominant weight.\cite{stephan} The adopted $\theta_{M}$ is consistent with this previous study.

The part $(1-P)H(t)P$ of the original extended Hubbard Hamiltonian neglected in the charge model changes the number of H-D pairs. Using strong-coupling perturbation theory, it has been shown that the parts that change the number of H-D pairs are the first order in the small parameter $T/(U-V)$.\cite{SCSOC1,stephan} Furthermore, $(1-P)H(t)P$ originates from the contribution of the components with a triplet-spin pair. Therefore, the neglected part $(1-P)H(t)P$ is the first order of $T/(U-V){\sqrt {1-c_{\rm S}^2(M)}}$.
For the parameters used in this paper, this quantity is as small as 0.11 at most to reproduce the optical conductivity of the Hubbard and extended Hubbard models (to be shown in Sec.~\ref{ssec:OCS}) even quantitatively.

In the following, we consider the linear absorption spectrum assuming a small vector potential, $A$.
We calculate exactly the optical conductivity in the charge model and in the 1D Hubbard and extended Hubbard models for a small cluster. A comparison of results is given in the following section. In a system with $N=4n+2$ ($N=4n$), with $n$ integer, the ground state of the 1D Hubbard model is a spin singlet (triplet). Since the spin-triplet state may affect the optical conductivity in the small-size system, we adopt a system size $N=14$ for a comparison. Furthermore, using the MBWFs, we also demonstrate in the following section
a newly developed approach to calculate the optical conductivity of strongly correlated electron systems 
of sufficiently large size, in which finite-size effects are negligible. In this method, the Hamiltonian matrix elements in the basis of MBWFs are obtained from the small cluster calculations that are then extrapolated to those for the larger systems.
From Eq.~(\ref{eq:thetaM}), $\theta_{M}$ for a system with $N=4n$ and that with $N=4n+2$ differ by $\pi$.
For the extrapolation, we adopt those $\theta_{M}$ for a system with $N=4n+2$ as well as a system with $N=4n$, because we are interested in optical excitations in the spin-singlet ground state. Note that the lowest-energy spin-singlet state is almost degenerate with that for the spin-triplet state for the one-dimensional Hubbard and extended Hubbard models. 
Furthermore, because the matrix elements for larger systems are needed as initial data for the extrapolation, we adopted a maximum system size of $N=16$ where exact diagonalization can be done practically. A twist in the boundary condition is introduced by adopting different $\theta_{M}$. The effects of the twisted boundary condition on optical properties are of order $1/N$, and they are negligible in the limit $N \to \infty$.

\section{Results} \label{sec:Results}

\subsection{Exact treatment} \label{ssec:OCS}
First, to validate our new model, we compare the optical conductivity spectra calculated using the charge model with that calculated 
using the 1D Hubbard and extended Hubbard models.

We use the translational symmetry and confine our argument to the zero center-of-gravity momentum frame.
Under these circumstances, the dimension of the Hilbert space of the charge model for $N=14$ is 44046, which is about 20 times smaller than that of the original models. 
Unfortunately, however, the computational limit for $N$ is about 26 even though the reduction of the dimension becomes more significant as $N$ increases.

To treat a larger system, we restrict the maximum number of H-D pairs, 
$M_{\rm max}$, where $1 \le M_{\rm max} \le N/2$.
The effect of this restriction on the spectra is discussed in this subsection.
However, even introducing this restriction, the practical upper limit of $N$ is 40, which is not sufficient to determine an overall spectral 
shape in the thermodynamic limit. 
In this subsection, the system size $N$ is fixed at 14 to perform an exact diagonalization of the Hubbard and extended Hubbard Hamiltonians. 
A further extension of the system size is discussed in the next subsection.

Under linear response, the electron--field coupling part of the Hamiltonian $H_{e-A}(t)$ is given by the first-order perturbation;
\begin{equation}
  H_{e-A}(t) = -A(t)\hat{J},
\end{equation}
where $\hat{J}$ is the current operator defined as
\begin{equation}
  \hat{J} = iT\sum_{n,\sigma}(c_{n,\sigma}^{\dagger}c_{n+1,\sigma}-\rm{H.c.}).
\end{equation}

In both the extended Hubbard and the charge models,
 we calculate the optical conductivity spectrum, which follows from the definition,
\begin{equation}\label{eq:OCS_def}
\sigma(\omega)={\gamma\over \omega N}\sum_\mu |\langle \Phi_{\mu} |\hat{J} |g\rangle |^2 
{1\over (\omega-E_\mu+E_g)^2+\gamma^2} \;,
\end{equation}
where $|g\rangle$ and $E_g$ are the ground state and ground state energy, respectively, and $|\Phi_{\mu}\rangle$
is the energy eigenstate associated with energy eigenvalue $E_\mu$. Here, the artificial broadening $\gamma$ is set to $0.1T$.

\begin{figure}[thbp]\centering
  \includegraphics[width=80mm]{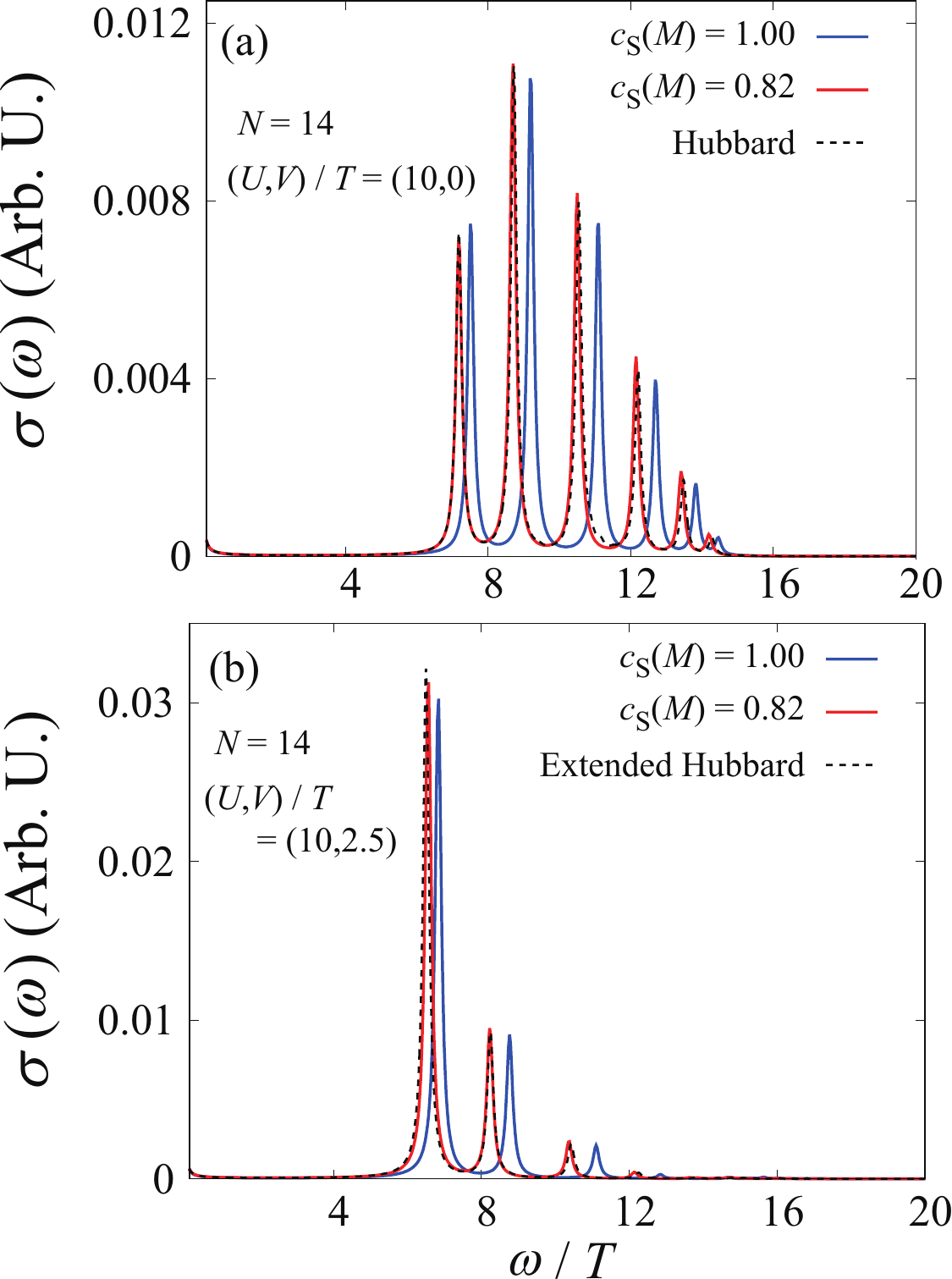}
  \caption{Comparison of the optical conductivity spectra of the Hubbard and extended Hubbard models
      and charge model for (a) $(U,V)/T=(10,0)$ and (b) $(U,V)/T=(10,2.5)$, and $N=14$. 
      All spectra are normalized so that $\langle g|\hat{J}^\dagger\hat{J}|g \rangle=1$ to compare the spectra for different $c_{\rm S}$ values, 
      and $M_{\rm max}=N/2$ holds.}
  \label{fig:OCS_cs_N14}
\end{figure}

In Fig.~\ref{fig:OCS_cs_N14}(a), the optical conductivity spectra for $(U,V)/T=(10,0)$ are shown.
Here, we do not restrict the maximum number of H-D pairs ($M_{\rm max}=N/2$). 
As mentioned in the previous section, an H-D pair is created only from a nearest-neighbor singlet spin pair and vice versa by virtue of the transfer term. The contributions of triplet spin pairs are included through the reduction factor $c_{\rm S}(M)$ in the charge model; see Eq.~(\ref{eq:OL}).
If the spin wave function of the charge model $f^{(M)}$ is approximated by the ground state of the $(N-2M)$ site 1D Heisenberg Hamiltonian, $c_{\rm S}(M)=0.82$ holds.~\cite{OLHG} 
From Fig.~\ref{fig:OCS_cs_N14}(a), the spectrum for the charge model with $c_{\rm S}(M)=0.82$ (red solid) is in good agreement with that for the Hubbard model (black dotted), justifying our assumption.
Furthermore, the good agreement shows that the spin-charge separation holds quite nicely.
In contrast, all the peaks of the charge model with $c_{\rm S}(M)=1$ for all $M$ (green solid) are blue-shifted about $0.5T$ in comparison with the Hubbard model.
This shift shows that the ground state is stabilized more largely than the optically excited states due to the overestimation of the transfer matrix elements related to creation and annihilation of an H-D pair.

The optical conductivity spectra for $(U,V)/T=(10,2.5)$ are also shown in Fig.~\ref{fig:OCS_cs_N14}(b).
The spectrum of the charge model with $c_{\rm S}(M)=0.82$ is again in good agreement with that of the exact spectrum.
The spectral features of the charge model with $c_{\rm S}(M)=1$ is almost the same as for $(U,V)/T=(10,0)$; that is, all peaks are blue-shifted about $0.5T$.

We therefore conclude that the charge model with $c_{\rm S}(M)=0.82$ is an effective model of the Hubbard and extended Hubbard models 
to investigate linear optical properties in small size clusters. Hence, $c_{\rm S}(M)$ is set to 0.82 from hereon. 
Although the contributions of the spin-triplet components are non-negligible, their effects are properly considered by the renormalization 
of the value of $c_{\rm S}(M)$.

\begin{figure}[thbp]
  \includegraphics[width=80mm]{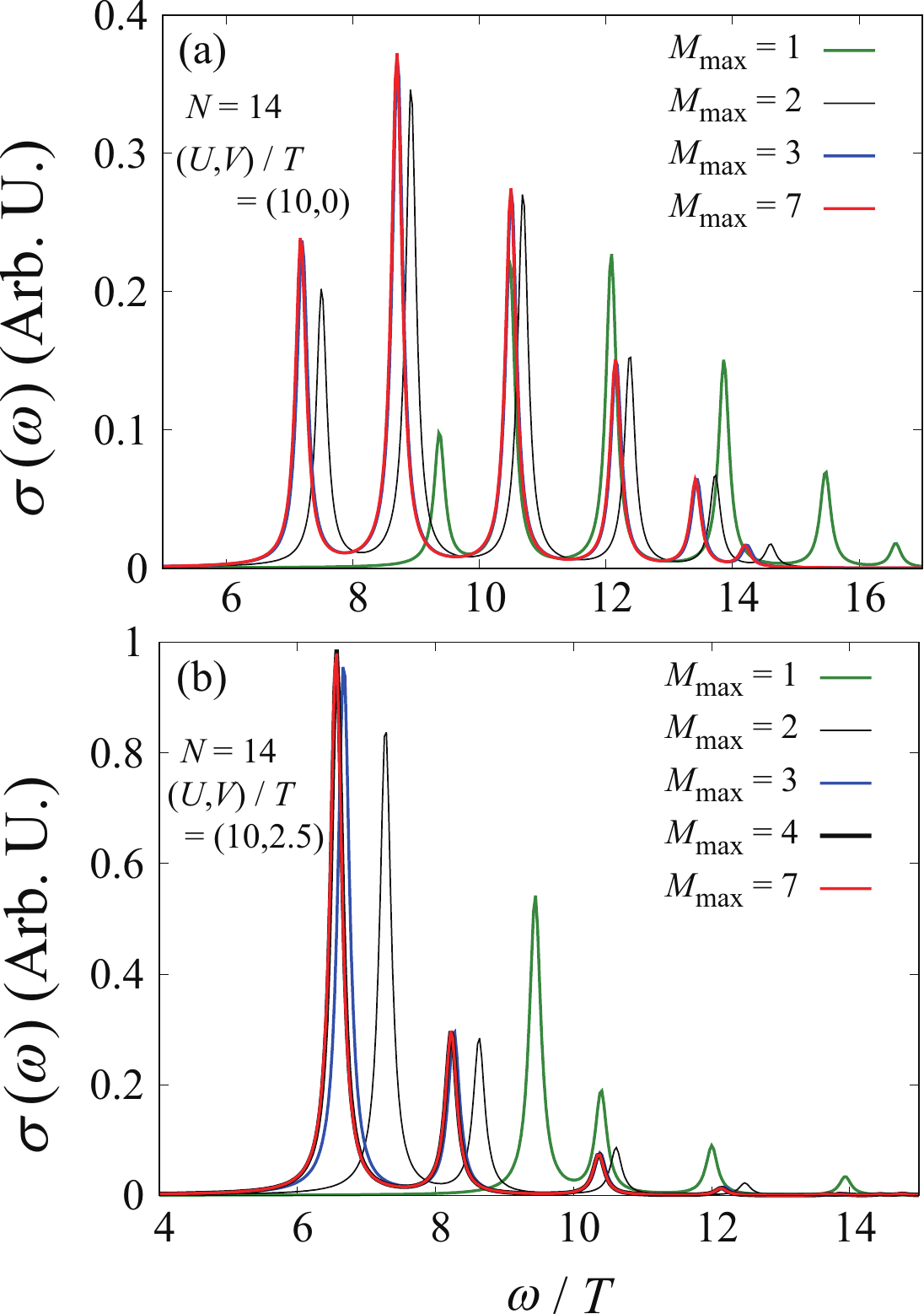}
  \caption{$M_{\rm max}$ dependence of the optical conductivity spectra for (a) $(U,V)/T=(10,0)$ and (b) $(U,V)/T=(10,2.5)$, and $N=14$. The red ($M_{\rm max} = 7$) and blue ($M_{\rm max} = 3$) lines in (a) and the red ($M_{\rm max} = 7$) and black thick ($M_{\rm max} = 4$) lines in (b) are indistinguishable.
      }
  \label{fig:OCS_nhd_N14}
\end{figure}

Next, we show the convergence of the optical conductivity spectra of the charge model in terms of $M_{\rm max}$. 
In Fig.~\ref{fig:OCS_nhd_N14}, the spectra of the charge model for several $M_{\rm max}$ values are shown. 
The spectra for $M_{\rm max}=1$ and $2$ are apparently blue-shifted in comparison with that for $M_{\rm max}=7$.
This shows that charge fluctuations (fluctuations in the number $M$ of H-D pairs) are significant and one- and two-H-D pair basis states cannot be enough to stabilize optically excited states.
We confirmed the numerical convergence of the spectra at $M_{\rm max}=3$ and $4$ for $(U,V)/T=(10,0)$ and $(10,2.5)$, respectively; the red ($M_{\rm max} = 7$) and blue ($M_{\rm max} = 3$) lines in Fig.~\ref{fig:OCS_nhd_N14}(a) and the red ($M_{\rm max} = 7$) and black thick ($M_{\rm max} = 4$) lines in Fig.~\ref{fig:OCS_nhd_N14}(b) strongly coincide.

A larger $M_{\rm max}$ value is required for $V>0$ than for the $V=0$ 
to describe optically excited states of $H^{\rm (C)}$ accurately. 
The explanation is that, because the nearest-neighbor H-D pairs are more stable for $V>0$ than for $V=0$ (the energy of formation is roughly given by $U-V$), 
multiple H-D pairs are created more easily in the former circumstance than in the latter. 
These results clearly show that charge fluctuations play an essential role with realistic $U\ (\sim 10T)$.

We also calculated the spectra for the relatively large system size, $N=40$, where the practical limit $M_{\rm max}$ is $5$.
 With $(U,V)/T=(10,0)$, the spectrum converges at $M_{\rm max}=5$.
 Although the spectrum for $(U,V)/T=(10,2.5)$ still does not converge completely for $M_{\rm max}=5$, 
 the largest peak position differs only by $0.2T$ from that for the extended Hubbard model calculated 
 by the time-dependent DMRG (t-DMRG) method; to be shown in the next subsection. See Appendix~\ref{app2} for the explanation of t-DMRG.
 We therefore use these results as a benchmark in the next subsection.
 
\begin{figure}[thbp]
  \includegraphics[width=80mm]{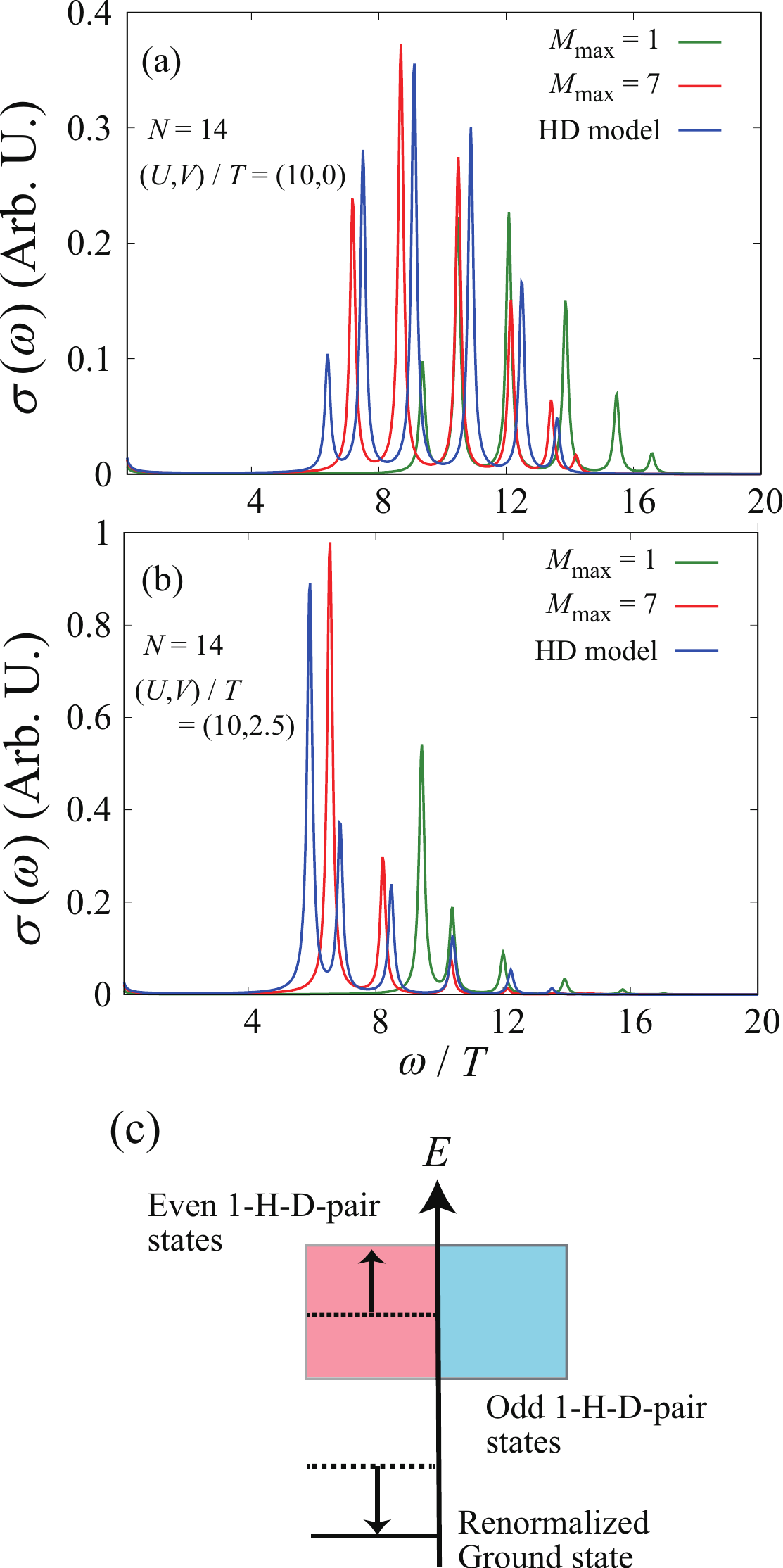}
  \caption{Comparison of the optical conductivity spectra of the charge model and HD model for (a) $(U,V)/T=(10,0)$ and (b) $(U,V)/T=(10,2.5)$, and $N=14$. (c) Schematic energy diagram for $M_{\rm max}=1$.}
  \label{fig:OCS_hdmodel_N14}
\end{figure}
 We mention here the difference between the charge model and the so-called holon-doublon (HD) model.\cite{NLOS2} 
 The HD model is an effective two-particle model, where the transfer of the H and D, as well as their Coulomb interaction are included. The essential difference between these two models is that the number of H-D pairs is set to one, and the annihilation and creation of H-D pairs do not occur in the HD model.
The ground state is stabilized by the charge fluctuation from the annihilation and creation of an H-D pair in the charge model.
Comparison of the optical conductivity spectra for $N=14$ calculated from the charge model and HD model are shown in Fig.~\ref{fig:OCS_hdmodel_N14}(a) and (b). We found that the center-of-gravity of the spectra calculated using the HD model shows better agreement with that calculated using the charge model with $M_{\rm max}=7$ than that calculated using the charge model with $M_{\rm max}=1$ for both $V=0$ and $V>0$ instances.
This feature is easily understood as arising from the difference in the stabilization of the two involved states, specifically, the ground state and the one-H-D-pair basis states with odd parity.
The ground state is the stabilized even for $M_{\rm max}=1$ in the charge model,
because the couplings between it and the even one H-D-pair basis states work there as seen in Fig.~\ref{fig:OCS_hdmodel_N14}(c).  
In more detail, the ground state couples with one of the even states most strongly.
As a result of this, the center of gravity of the whole even states does not change largely.
The odd one-H-D-pair states are, in contrast, not stabilized in the absence of couplings with multiple H-D-pair basis states, as arises for $M_{\rm max}=1$.  
This imbalance yields an incorrect large gap in the spectrum of $M_{\rm max}=1$.
Meanwhile, the cancellation of inaccuracies results in the better optical gap for the HD model than for the charge model with $M_{\rm max}=1$ by chance.
However, the HD model cannot reproduce the detailed distribution of spectral peaks of the charge model with $M_{\rm max}=7$.
Furthermore, the HD model gives incorrect optical gaps for smaller $U\ (\sim 5T)$ due to the neglection of charge fluctuations, which will be shown in Fig.~\ref{MBWF-6}.


\subsection{Many-Body Wannier Functions} \label{ssec:MBWFs}
In the preceding sections, we introduced the charge model and demonstrated the optical conductivity spectra calculated 
using the model with a small system size that can be treated exactly. 
Even introducing a restriction to the maximum number of H-D pairs, the practical upper limit is $N=40$, 
which is not sufficient to determine an overall spectral shape in the thermodynamic limit.
We, therefore, try the calculation for much larger system sizes and present the spectra in those cases based on the newly developed 
many-body Wannier functions.

For conventional Wannier functions, the full Bloch functions constitute a complete orthogonal set
for the one-body states. The Wannier functions are obtained from the former using a unitary transformation to make the latter
as localized as possible.
The benefit of these functions is the direct descriptions of the nature of the corresponding band dispersion,
which is independent of the assumed system size. They are used to estimate the model parameters such as transfer energy and on-site
repulsion energy. The resultant models are well-known to play substantial roles in the investigation of much more 
subtle aspects such as electron correlations beyond one-body treatments.

We apply this ``philosophy" of the Wannier functions to the present charge model. What is essential is hence
the  construction of the many-body counterpart, which is defined locally, being almost free from the system size.
Because of this local nature, they provide a practically useful basis set in the many-body problem.
More specifically,  we define a subspace of important many-body states as a complete orthonormal set and transform it into 
another complete orthonormal set of which the states are spatially localized in a pre-defined meaning.

These MBWFs have several advantages when compared with the other methods to determine an overall spectral shape in the thermodynamic limit.
First, the dynamical DMRG (DDMRG) method is known to provide numerically almost
exact results.~\cite{DDMRG,DDMRG2,DDMRG3,DDMRG4,DDMRG5,DDMRG6,DDMRG7,DDMRG8}  
In spite of its accuracy, we often experience 
difficulty in knowing the nature of a spectral feature.  This difficulty arises from the repeated basis transformations performed in the DDMRG.
In contrast, the basis transformation in the MBWF method is performed only once, and we easily translate a result into that based 
on the  original bare basis states.  Second, the technique of the quantum Monte Carlo (MC) is one of the non-perturbative methods. 
In some cases, it gives reliable spectral results,~\cite{QMC1,QMC2}
although the problem of analytical continuation still requires careful treatment.  In addition to this demerit, the difficulty in knowing the nature
of a spectral feature also applies to this method.  Lastly, analytical methods are also compared with the method of MBWFs.  
The method based on the Bethe ansatz leads to an analysis in the large-$U/T$ limit,~\cite{stephan}
while a field-theoretical method is limited to the small-$U/T$ region.~\cite{DDMRG,DDMRG2}
As will be shown in this subsection, the method of MBWFs has a wide application range with intermediate and strong $U/T$ values.
\begin{figure}[tb]
	\begin{centering}
	\includegraphics[width=0.65\textwidth]{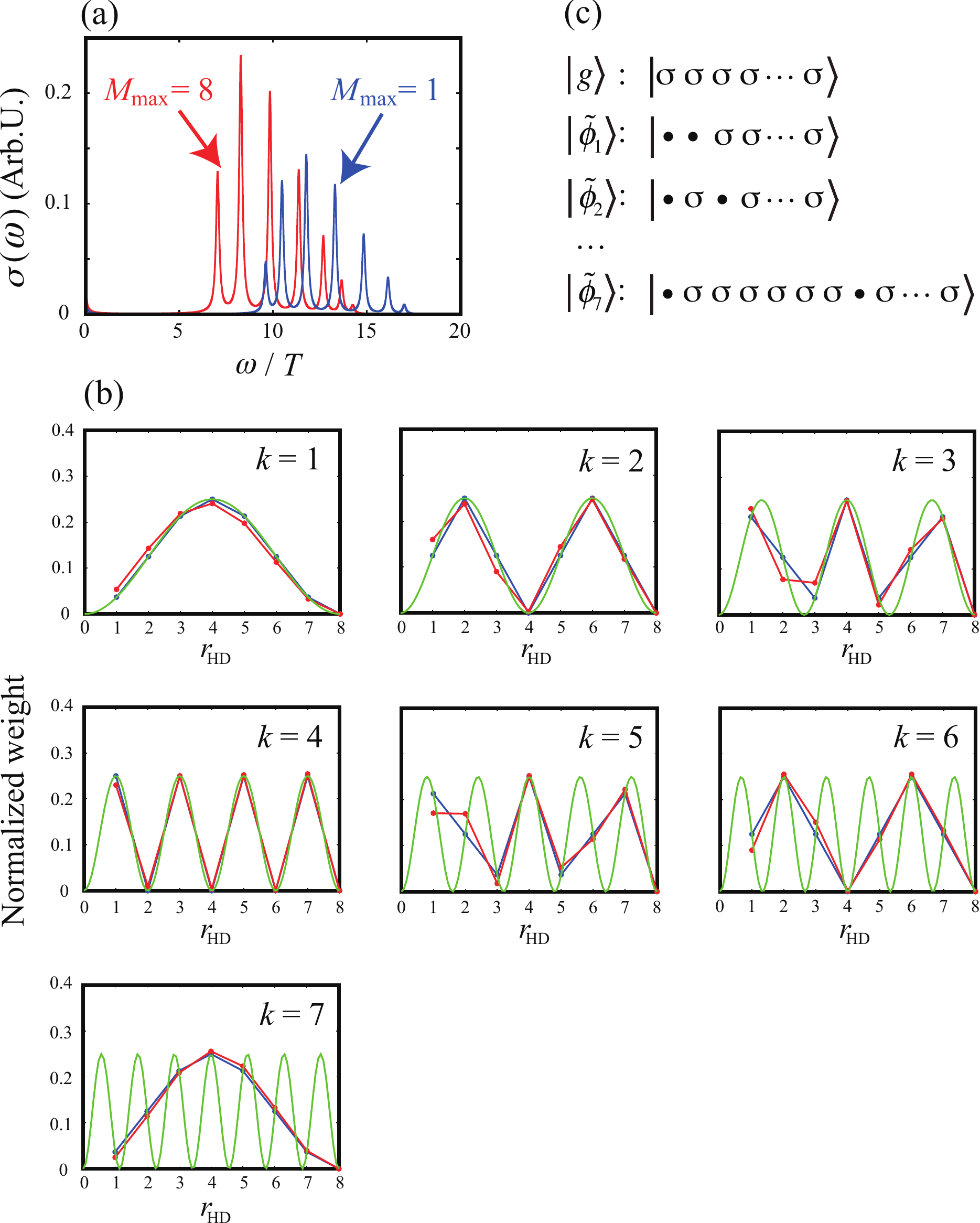}
	\caption{(a) Optical conductivity spectra for $(U, V)/T=(10, 0)$ and $N$=16, 
	with $M_{\rm max}=1$ (blue line) and $M_{\rm max}=8$ (red line);
	The ground state, which is repelled  by the
	even one-H-D-pair basis state with the H-D distance as one, is stabilized to a lower energy;
	(b) Normalized weight distributions as a function of $r_{\rm HD}$. The blue (red) lines mark those for
	 $M_{\rm max}=1$ ($M_{\rm max}=8$). The green lines obey the sinusoidal functions defined in the text;
	 (c) Schematics for the site-localized basis states for $M_{\rm max}$=1.  For all the basis states, an odd H-D state is assumed.  
	 For example, $|\bullet\bullet\rangle$ is $(|{\rm HD}\rangle-|{\rm DH}\rangle)/\sqrt{2}$, 
	 and $|\sigma\rangle$ means a site with an unpaired spin.}
		\label{MBWF-1}
	\end{centering}
\end{figure}

Before entering into the actual construction of the MBWFs,
we discuss in more detail the optical conductivity spectrum for $H^{\rm (C)}$ in a small cluster, 
specifically to know the nature of each peak.
Here, we use a system with $N=16$ as a starting point of our construction.  In principle, the initial system
size is required to be sufficiently large to contain the spatial extension of MBWFs.  
For the present cases, we found that the choice of 16 sites is considered 
to be adequate.  Furthermore, we use the translational symmetry and restrict our argument within the frame of zero 
center-of-gravity momentum.
In Fig.~\ref{MBWF-1} (a), the spectrum of the optical conductivity calculated with artificial broadening $\gamma=0.1T$ is shown.  
The number of H-D pairs has no restriction for the solid red curve, which means that the maximum number of H-D pairs, $M_{\rm max}$, is 8. 
In contrast, when $M_{\rm max}$ is set to one (blue dotted curve), then only the bare ground state, namely, the charge vacuum, and one-H-D-pair basis states
are included.   When we compare the two spectra, the apparent difference is the larger optical gap in the latter, which is the same feature as seen in the Fig.~\ref{fig:OCS_hdmodel_N14}(a).

In Fig.~\ref{MBWF-1}(b), the H-D distance ($r_{\rm HD}$) distributions are
shown for each eigenstate corresponding to the seven principal peaks in the spectra [Fig.~\ref{MBWF-1}(a)].  
Here, all the states are parity-odd, and the numbering is in increasing order of the eigenenergy.
Note that the sampling for $M_{\rm max}=8$ is performed with respect to 
the one-H-D-pair states and that the summation for all the distances is normalized to unity.  First, the curves for $M_{\rm max}=1$ obey
the exact functions; that is, ${1\over 4}\sin^2({\pi\over (N/2)}kr_{\rm HD})$ ($1\leq k\leq (N/2-1)$), as expected from their 
unperturbed nature as Bloch states, whereas those for $M_{\rm max}=8$ are slightly deformed from those.  We emphasize that this does not mean necessarily
weak renormalization because of  multiple pair excitations, as the ratio of one-H-D-pair basis states within the whole weight is largely reduced to
 70\%--75\% for $M_{\rm max}=8$, in contrast to 100\% for $M_{\rm max}=1$.

We next construct the MBWFs given the above eigenstates. For $M_{\rm max}=1$, this is trivial.  Specifically, defining
$\{|\tilde{\phi}_k\rangle \}$ as the odd one-H-D-pair basis state having the H-D distance of $k$ $(k=1,2,...,7)$, such states constitute
a complete orthogonal set for the optically active states.  In other words, they provide the whole transitions moments 
from the ground state, which assures that the optical conductivity spectrum is described exactly by this subspace, that is, 
the above excited states and the ground state.  
In Fig.~\ref{MBWF-1}(c), we illustrate some of the  actual $|\tilde{\phi}_k\rangle$'s.
Note that the basis state with distance $k=8$ is parity-even. 
The Bloch states corresponding to the seven principal peaks defined as $\{|\phi_k\rangle \}$ 
are exactly expressed as  $| \phi_k \rangle=\sum_{k'} V^{({\rm tr})}_{kk'} |\tilde{\phi}_{k'}\rangle$, where $V^{({\rm tr})}$ is a unitary matrix.
By a simple analysis, we determine the unitary matrix to be
\begin{equation}
V^{({\rm tr})}_{kk'}={2\over \sqrt{N}} \sin({2\pi \over N}kk') \;\;  (1\le k, k' \le 7) \; .
\end{equation}
From these equations, we now see that the functions $\{ |\tilde{\phi}_{k'}\rangle \}$ play the roles of MBWFs.  

For $M_{\rm max}=8$, we again choose seven principal peaks.  They dominate the whole transitions moments
and we expect the optical conductivity spectrum to be described by the ground state and the seven
corresponding energy eigenstates $\{|\phi_k\rangle\}$ very accurately. In this case, we try a reverse transformation as
\begin{equation}
| \tilde{\phi}_k \rangle=\sum_{k'} V^{({\rm tr})}_{kk'} |\phi_{k'}\rangle \; ,
\end{equation}
using the same matrix $V^{({\rm tr})}$, because the behavior of the Bloch states confined in the one-H-D pair basis states
is similar to that for $M_{\rm max}=1$; see Fig.~\ref{MBWF-1}(b).  We emphasize that $|\tilde{\phi}_k \rangle$ 
is expressed as a linear combination of many basis states, which are separated into one-H-D pair basis states and multi-H-D pair basis states.
In this transformation, the former part is {\it localized}, in the meaning that the one-H-D pair basis with the
H-D distance being $k$ has a relative weight more than 94\% among all the one-H-D pair bases.
In this sense, we regard them as MBWFs.  Meanwhile, the latter part, i. e., the part composed by
the multi-H-D pair basis states, is regarded as a non-trivial fluctuation associated with this MBWF and plays an essential role in the determination of
the effective model below.

\begin{figure}[tb]
	\begin{centering}
	\includegraphics[width=0.85\textwidth]{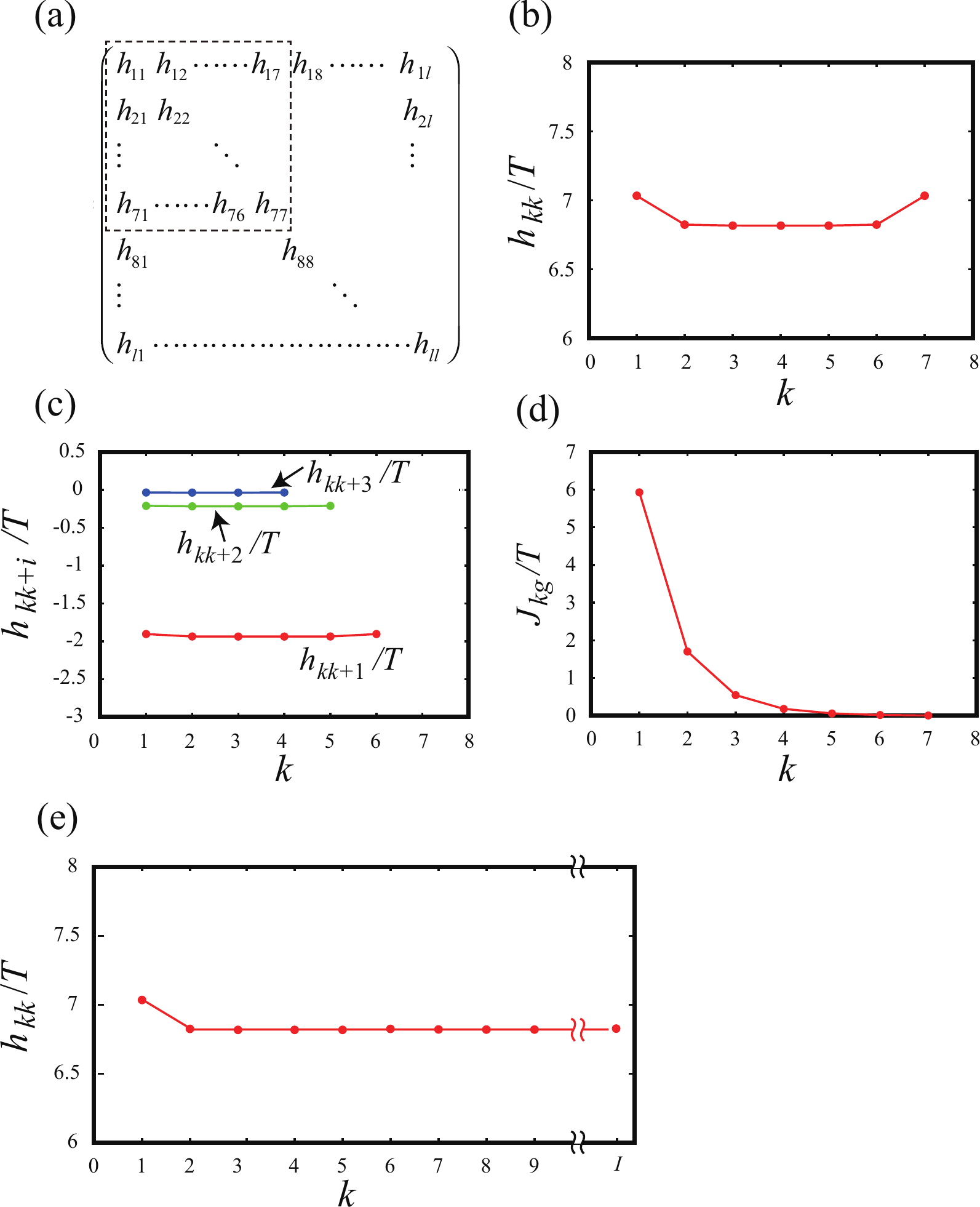}
	\caption{(a) Structure of the matrix elements calculated using MBWFs (the part within the dotted square) 
	        and their extensions (outside the square). 
	       Calculated (b) diagonal and (c) off-diagonal matrix elements of $H^{\rm (C)}$. (d) Calculated matrix elements of the current operator. 
	        (e) An example of extrapolation corresponding to (b).}
		\label{MBWF-3}
	\end{centering}
\end{figure}

The obtained MBWFs are used to evaluate the matrix elements of $H^{\rm (C)}$.  In Fig.~\ref{MBWF-3}, we show the matrix
elements, $h_{kk'}$, which are defined as $\langle \tilde{\phi}_k | H^{\rm (C)} |\tilde{\phi}_{k'} \rangle$ with $k$ and $k'$ being 1$\sim$7, 
as specified within the dotted square of Fig.~\ref{MBWF-3}(a).
In each of Fig.~\ref{MBWF-3} (b)--(d), we plot the matrix elements along the diagonal lines.  In Fig.~\ref{MBWF-3}(b), all the elements are
diagonal elements and are almost constant except for the slightly larger values at the boundaries, i.e., $k=1$ and 7.
Among the off-diagonal elements, $h_{kk+1}$ in Fig.~\ref{MBWF-3}(c) take large values near to $-2T$.  
For $M_{\rm max}=1$, the corresponding values are exactly $-2T$, which represents twice the transfer energy 
because the HD distance changes with the movements of both H and D.
The values
close to $-2T$ are surprising because the states are substantially renormalized due to multiple excitations, as already mentioned.
We attribute this peculiar property to a coherent build up of the matrix elements in each subspace of the $M$ H-D pairs.
Regarding the remaining off-diagonal elements, we consider those up to $h_{kk+3}$ plotted in Fig.~\ref{MBWF-3}(c).
As expected from their trends, the elements such as $h_{kk+i}$ with $i\ge4$ are very small, and we neglect them in the
following calculation.  The effective model, $h_{\rm eff}$, is defined using $h_{kk'}$ with $k'\le k+3$ and their transposed elements.
Note that elements $h_{kk+i}$ with $i\ge2$ vanish completely for $M_{\rm max}=1$,  representing the
short-range nature of the charge model,.

Figure \ref{MBWF-3}(d) plots the matrix element of the current operator, $J_{kg}$, 
which is defined as $\langle \tilde{\phi}_k | \hat{J} |g\rangle$, with $|g\rangle$ being the ground state for the 16 sites.
We again remark that the  same quantities vanish except at $r_{\rm HD}=1$, for $M_{\rm max}=1$. 
Owing to multiple excitations, the element for $M_{\rm max}=8$ is no longer localized at $k=1$; instead, they decay smoothly at longer distances. 
Here, it is crucial for the MBWF scheme that this decay is contained within the system size.
In this regard, we find no serious problem for the present and other parameter sets used in this article.
  
We next enlarge the obtained effective Hamiltonian by extrapolating the matrix elements [Fig.~\ref{MBWF-3}(a)]. 
Before entering into the details, we explain the basic strategy of our extrapolation.  In particular, 
we focus on the extrapolation of the optical conductivity spectrum.   Although the extrapolation of the ground state 
itself will be an issue in other studies, the purpose of this study is to predict the optical spectrum 
in large systems.  For this reason, we focus on the excitation energies and redefine the effective model 
as $\tilde{h}_{\rm eff} \equiv h_{\rm eff}-E_g$, subtracting the ground state energy for the 16 sites.
By this substitution, we can determine the spectrum efficiently without finding the ground state in large systems directly.
The actual extrapolation for $r_{\rm HD}\ge 8$ is
rather straightforward for the matrix elements of $\tilde{h}_{kk+2}$ and $\tilde{h}_{kk+3}$, that is, approximating all of them as
the averaged values in the present system  size.  The matrix elements for the current operator are also extrapolated
straightforwardly, that is, padding the elements of $r_{\rm HD}\ge 8$ with zeroes and multiplying them by $\sqrt{N_{\rm ex}/N}$.
Here, the enlarged system size is $N_{\rm ex}$, which is expressed as $N_{\rm ex}=2I+2$, with $I$ being 
the maximum H-D distance of odd one-H-D-pair basis states.
This factor is required because the matrix element, $J_{kg}$, is proportional to the square root of the 
system size in the thermodynamic limit.
The diagonal elements, by contrast, needs some care.  As we have already mentioned,
the diagonal elements take slightly larger values at the boundaries.  Based on our inspection, the final results, i.e., the spectral shapes in the
enlarged systems, tend to depend on the boundary effect, particularly at $k=1$.  We therefore keep this boundary effect 
[Fig.~\ref{MBWF-3}(e)].
Meanwhile, we neglect  the boundary effect at the farthest point, i. e., at $r_{\rm HD}=I$, because its effect on the spectrum is negligible, 
as readily expected from the behavior of the $\hat{J}$ matrix elements.
\begin{figure}[tb]
	\begin{centering}
	\includegraphics[width=0.5\textwidth]{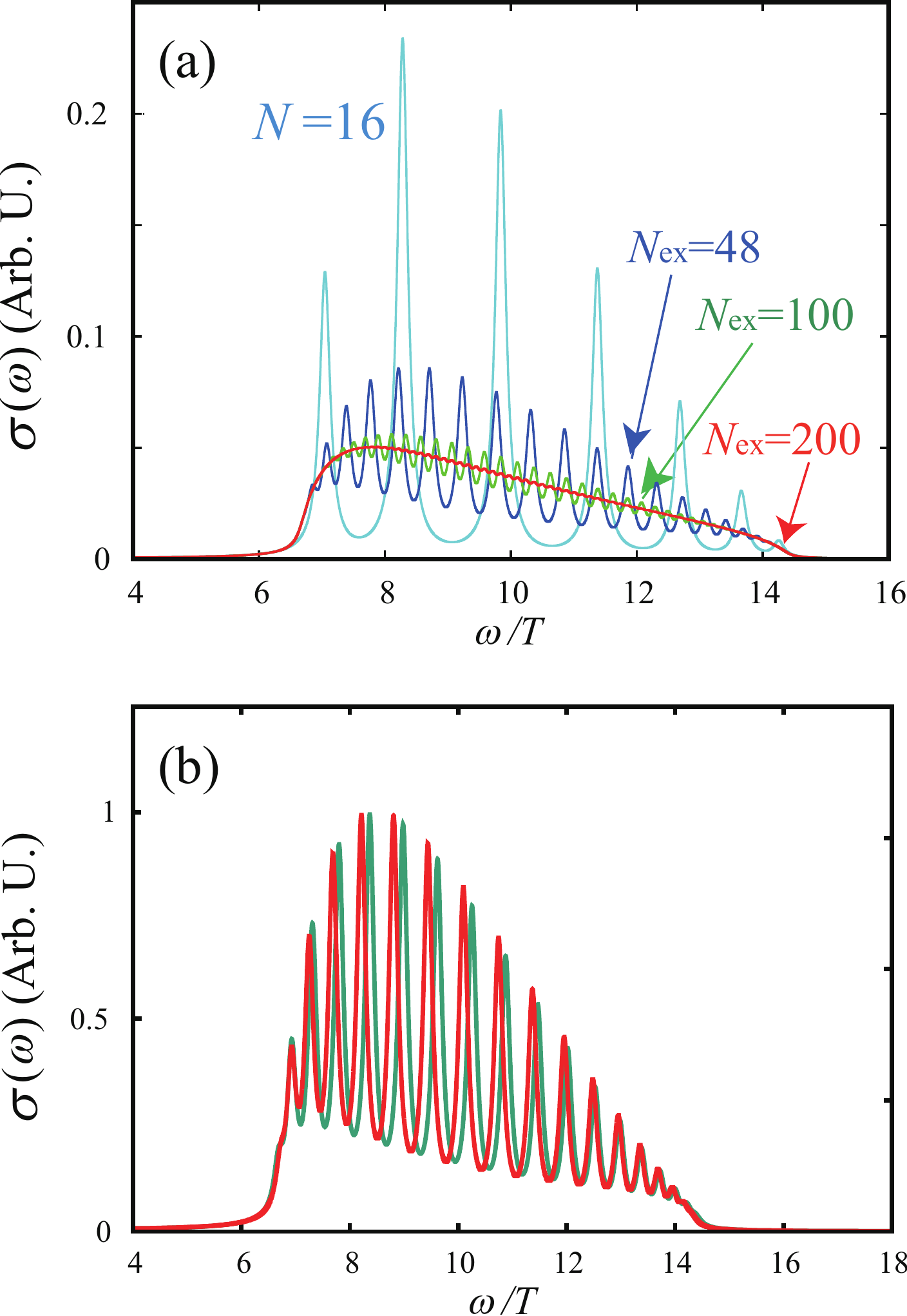}
	\caption{Optical conductivity spectra in the charge model with $(U, V)/T=(10, 0)$.  (a) Results obtained using MBWFs for several enlarged system sizes;
	(b) Comparison with the direct calculation at $N=40$.
	         The red and green lines represent respectively the result obtained using MBWFs and that by a direct calculation with truncation.
	         Each intensity plot is normalized by its maximum.
	       }
		\label{MBWF-4}
	\end{centering}
\end{figure}

Based on the enlarged effective model, we calculate the optical conductivity spectrum, which follows a slightly changed definition,
\begin{equation}\label{eq:MBWF}
\sigma(\omega)={\gamma\over \omega N_{\rm ex}}\sum_{\mu=1}^I  |\langle \Phi_{\mu} |\hat{J} |g\rangle |^2 
{1\over (\omega-E_\mu^{({\rm eff})})^2+\gamma^2} \;,
\end{equation}
where  $|\Phi_{\mu}\rangle $  and $E_\mu^{({\rm eff})}$
denote respectively the $\mu$-th eigenstate  and its eigenenergy of the enlarged effective model, $\tilde{h}_{\rm eff}$.
Using this definition, the spectra are calculated
for several $N_{\rm ex}$'s [Fig.~\ref{MBWF-4}(a)]. The spectral shape appears to have almost completely converged
with the system size around 200.  To confirm the validity of the present treatment of the MBWFs, 
we also show the result for  $N_{\rm ex}=40$ as well as  that by a direct diagonalization with truncation [Fig.~\ref{MBWF-4}(b)].
Note that this is the maximum size by which we calculate the spectra directly.  We emphasize that the present choice of
$N=40$ and $M_{\rm max}=5$ is considered to be balanced, because it gives an almost converged spectrum when we increase the latter
keeping the former fixed.  Comparing the two spectra, we conclude that they coincide with each other within a practical tolerance
and that the present treatment works satisfactorily at least for the present parameter set.

\begin{figure}[tb]
	\begin{centering}
	\includegraphics[width=0.4\textwidth]{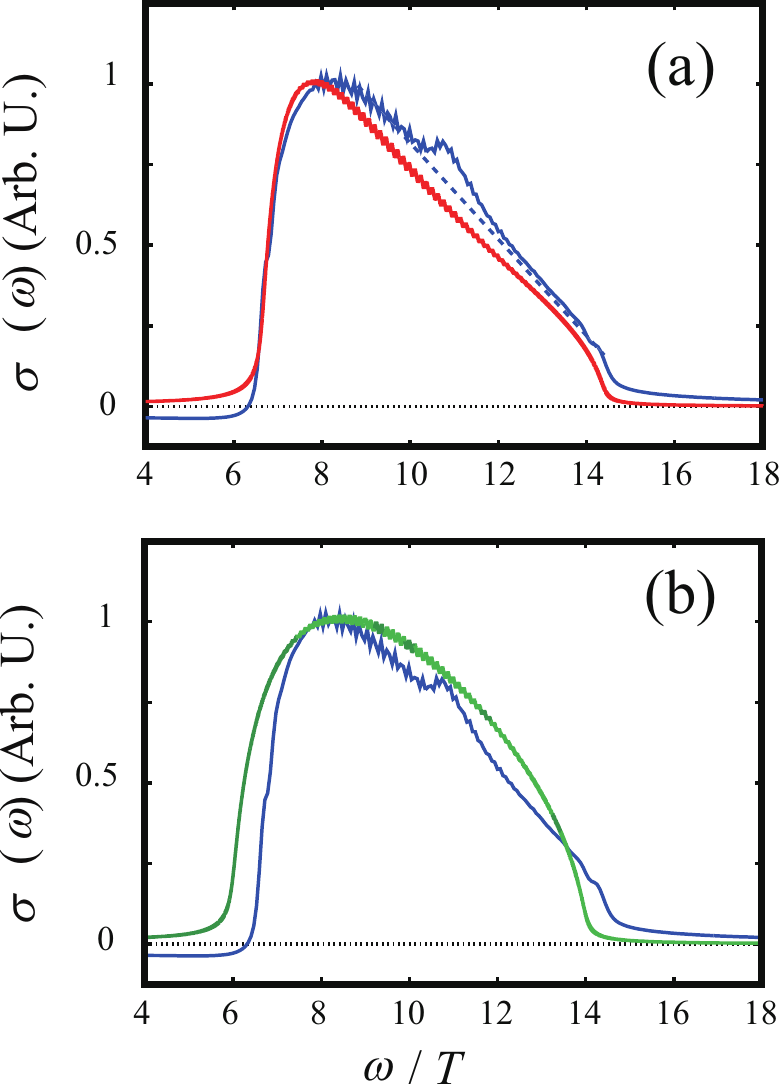}
	\caption{Comparison of optical conductivity spectra at $(U, V)/T=(10, 0)$.  In both the graphs, the blue line represents
	the result obtained using t-DMRG applied to the Hubbard model.  The red line in (a) represents 
	the spectrum obtained using MBWF for the charge model
	with $N_{\rm ex}=200$. The green line in (b) represents the spectrum obtained by the HD model 
	with $N=200$.  Each intensity is normalized by its maximum.}
		\label{MBWF-5}
	\end{centering}
\end{figure}
We next argue the significance of the charge model particularly compared with the conventional HD model.
In Fig.~\ref{MBWF-5}(a), we again show the spectrum for the charge model with $N_{\rm ex}=200$ and $(U, V)/T=(10, 0)$ (red line) as well as
that for the Hubbard model with the same parameter set.  Note that the latter spectrum is  obtained by t-DMRG for $N=80$ 
(blue line).  
Although we find a discrepancy on the high-energy side of the absorption band,
both the high and low energy edges are well reproduced.  Of note is a small hump seen around $\omega/T=11$ for the
t-DMRG result, which is associated with the spin degrees of freedom~\cite{DDMRG} and does not appear in the spectrum for the charge model.  
If we exclude this hump as shown by the dotted blue line, the discrepancy can be considered to be rather small.
In Fig.~\ref{MBWF-5}(b),  we compare the results obtained by the HD model (green line) and the t-DMRG.  
In the HD model, only the one-H-D-pair basis states are considered, whereas, in the charge model with $M_{\rm max}=1$, 
the  ground state is also included.  Although the latter is an extension of the former, this partial extension instead gives
an incorrect large optical gap, as already mentioned.  In this regard, the HD model gives a moderately incorrect
optical gap.  For instance, the optical gap is smaller by about $T$ from that determined by t-DMRG.
Furthermore, we find a discrepancy in the whole spectral shape.
Note also that the asymmetry in the HD
model comes only from the factor of $1/\omega$ included in the expression for conductivity.  The spectrum obtained by t-DMRG,
which is expected to be close to that of the charge model, is more asymmetric than that from the HD model,
indicating an appreciable amount of renormalization inherent in the spectrum.  
Regarding the nature of this renormalization, we believe  that multiple excitations
of the H-D pairs that we have already mentioned play an essential role.  Indeed, a comparison in Fig.~\ref{MBWF-5}(a) suggests that the spectrum
for the charge model reproduces the asymmetry existing in the spectrum for the Hubbard model although the asymmetry
seems to be slightly exaggerated in the former.

\begin{figure}
	\begin{centering}
	\includegraphics[width=0.5\textwidth]{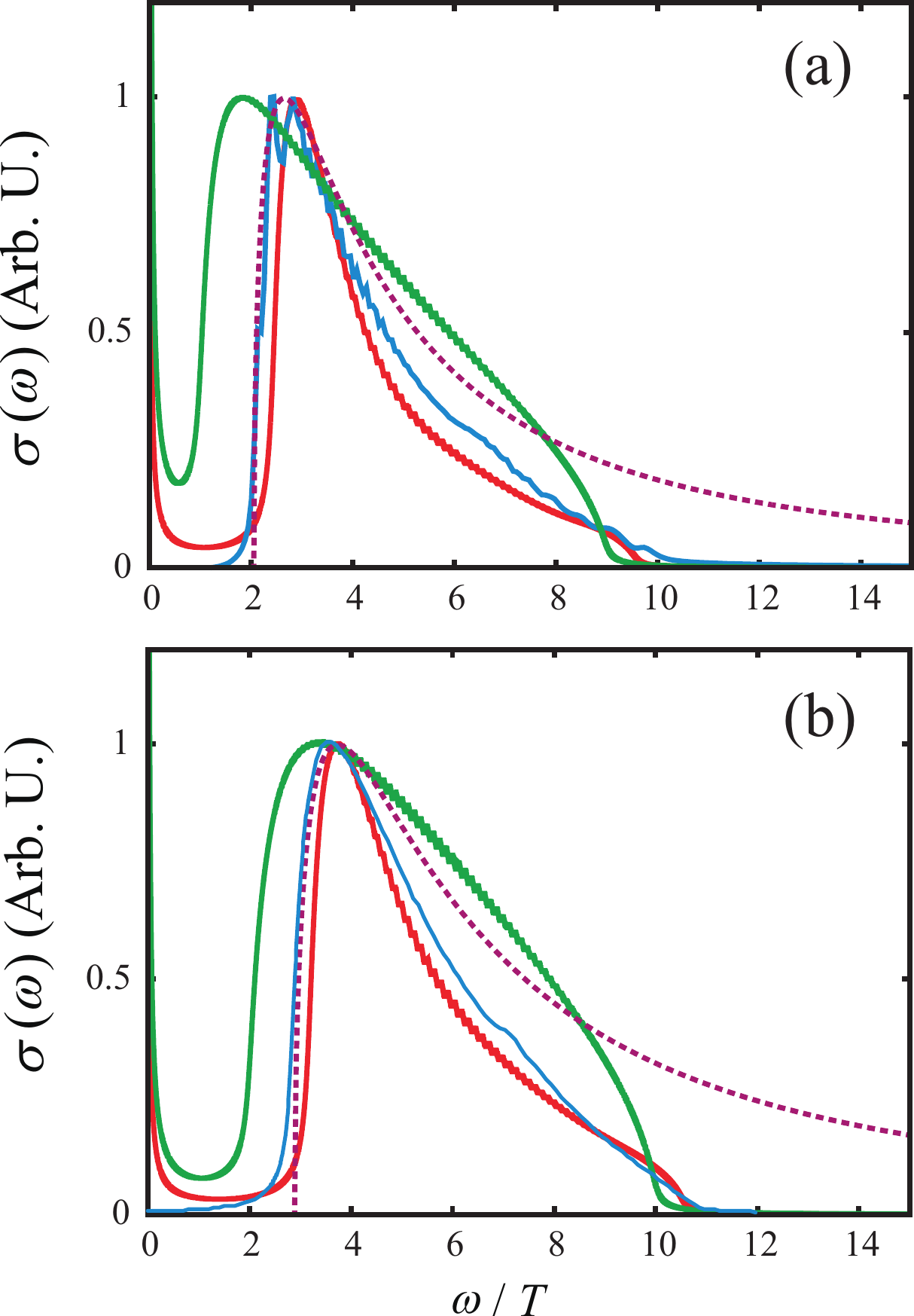}
	\caption{Optical conductivity spectra for (a) $(U, V)/T=(5, 0)$ and (b) $(U, V)/T=(6, 0)$. 
	 The red lines represent the results obtained using MBWFs based on the charge model with $N_{\rm ex}=200$.  
	 The blue lines in (a) and (b) represent the results obtained based on DMRG-derived methods applied for the Hubbard model, and we used
	 respectively t-DMRG with $N=80$ and DDMRG with $N=100$.  The results by the field theoretical method applied for the
	 Hubbard model are expressed by the dashed purple lines, while those by the HD model by the green lines.
	        Each intensity profile has been normalized by its maximum.
	        The DDMRG result is reprinted from Ref.~\onlinecite{DDMRG}. }
		\label{MBWF-6}
	\end{centering}
\end{figure}
From here on, we discuss the validity of the present method for smaller $U$ values.  To make the argument 
transparent, we confine the discussion to vanishing $V$.
Although it depends on the material, the actual $U$ values associated with molecular solids and metal oxides, in which 
strongly correlated electron systems appear, range very roughly from $U/T=5$ to 10 or much larger values.  
In this situation, we think that there are at least two crucial points regarding the validity of the present method.
One is the validity of the charge model itself.  Whereas the charge model in the absence of $V$
coincides with the exact theory of the Hubbard model in the limit of infinite $U$, 
situations with finite $U$ values should be checked by additional analyses.  
In this sense, the approximate coincidence seen in Fig.~\ref{MBWF-5}(a)
justifies the charge model at least for $U/T=10$, whereas a check for smaller $U$ values remains.
The other point is the validity of the MBWF.  The present MBWF describes a photoexcited state as a renormalized H-D pair state.
Although its extension is possible in principle, for instance, MBWFs for two pair states, at present, there are drawbacks for small $U$
because the nature of one pair state is gradually lost as $U$ decreases.  For this reason, we think that
checks are required of the results down to $U/T=5$.  
In Fig.~\ref{MBWF-6}, we show the spectra calculated for $U/T=5$ and 6 (red lines).
Here, they are compared with the results obtained using the Hubbard model with the corresponding parameters using t-DMRG
 [Fig.~\ref{MBWF-6}(a)] and the dynamical DMRG~\cite{DDMRG} [Fig.~\ref{MBWF-6}(b)].  
 In both cases, we see that the coincidences are satisfactory at least for the purpose of
determining the overall spectral shape.  On the basis of this result, we believe that the region in which the validity of the present
method is assured extends at least down to $U/T=5$.  
As added remarks,  we also show the results based on the other methods, that are,  the HD model and the field theoretical method\cite{DDMRG,DDMRG2} applied for the Hubbard model marked by the green line and the dashed purple line, respectively.
Regarding the HD model, the discrepancies of the results from those by the DMRG-derived methods are more conspicuous, as seen in the
red shift of the lower edge and the exaggerrated  feature on the high-energy side. The field theoretical method, on the other hand, reproduces
the correct position of the lower edge, although the high-energy side deviates largely from that of the DMRG-derived method. We note that
the spectrum for $U/T$=3 by the field theoretical method coincides almost satisfactorily with that by the DDMRG.~\cite{DDMRG}  Although
we do not show it explicitly, our method underestimates the tail structure on the high-energy side.

As a final topic in this section, we argue the case of finite $V$.  To consider the effect of $V$, we treat it as a perturbation.
We first determine the MBWFs for vanishing $V$ excluding the term $H^{\rm (C)}$ associated with
$V$ (hereafter called the $V$ term).  After that, 
we take the matrix element of the whole $H^{\rm (C)}$ including the $V$ term and diagonalize it.  This treatment is somewhat analogous
to the so-called single-configurational-interaction approximation, which also introduces the excitonic effect into the one-electron
excitations that are prepared appropriately. Here, we use the parameter set of $(U, V)/T=(10,2.5)$.
As the upper limit for $V$ with $U/T=10$ is almost 5 in the Mott-insulator phase, the present $V$ value is intermediate.
We avoid larger $V$ values, because the truncation of $M_{\rm max}=5$ used in the direct calculation becomes insufficient, and 
therefore, confirming the accuracy of the result with confidence is difficult.  The actual procedure is similar 
to the case of vanishing V.  Namely, we take the matrix elements considering the $V$ term using the MBWFs and extrapolate them, 
as described in detail in the Appendix~\ref{app3}.

In Fig.~\ref{MBWF-8}(a), the calculated spectrum for $N_{\rm ex}=40$ is shown with the result from the exact calculation for $N=40$, which
is obtained again with truncation of $M_{\rm max}=5$. We find good agreement in the spectral shapes, which justifies the treatment of
MBWFs even in the presence of $V$.  Figure~\ref{MBWF-8} (b) shows the result obtained using MBWFs for $N_{\rm ex}=200$ and a comparison with
the result obtained using t-DMRG applied to the extended Hubbard model with $N=80$ [Fig.~\ref{MBWF-8}(c)].  Here, the width of the artificial
broadening $\gamma$ is 0.1$T$ for the former, whereas it increases slightly to 0.14$T$ for the latter, 
to give almost the same main peak width.  
This is because a different definition of broadening is used in the formalism of t-DMRG, one that is not based on an expression like Eq.~(\ref{eq:MBWF}).  Because there is
no established way to convert the value of $\gamma$ at present, we have adjusted $\gamma$ of t-DMRG by fitting. This adjusted $\gamma$
provides almost the same peak width for the sharpest part of each spectrum.  Apart from this similarity, we also notice several other common features,
for example, the width of the high-energy tail and the position of the lower edge.  Based on this consistency, we also conclude that the
charge model is a good approximation to the extended Hubbard model from the viewpoint of optical conductivity spectra.
\begin{figure}[tb]
	\begin{centering}
	\includegraphics[width=0.5\textwidth]{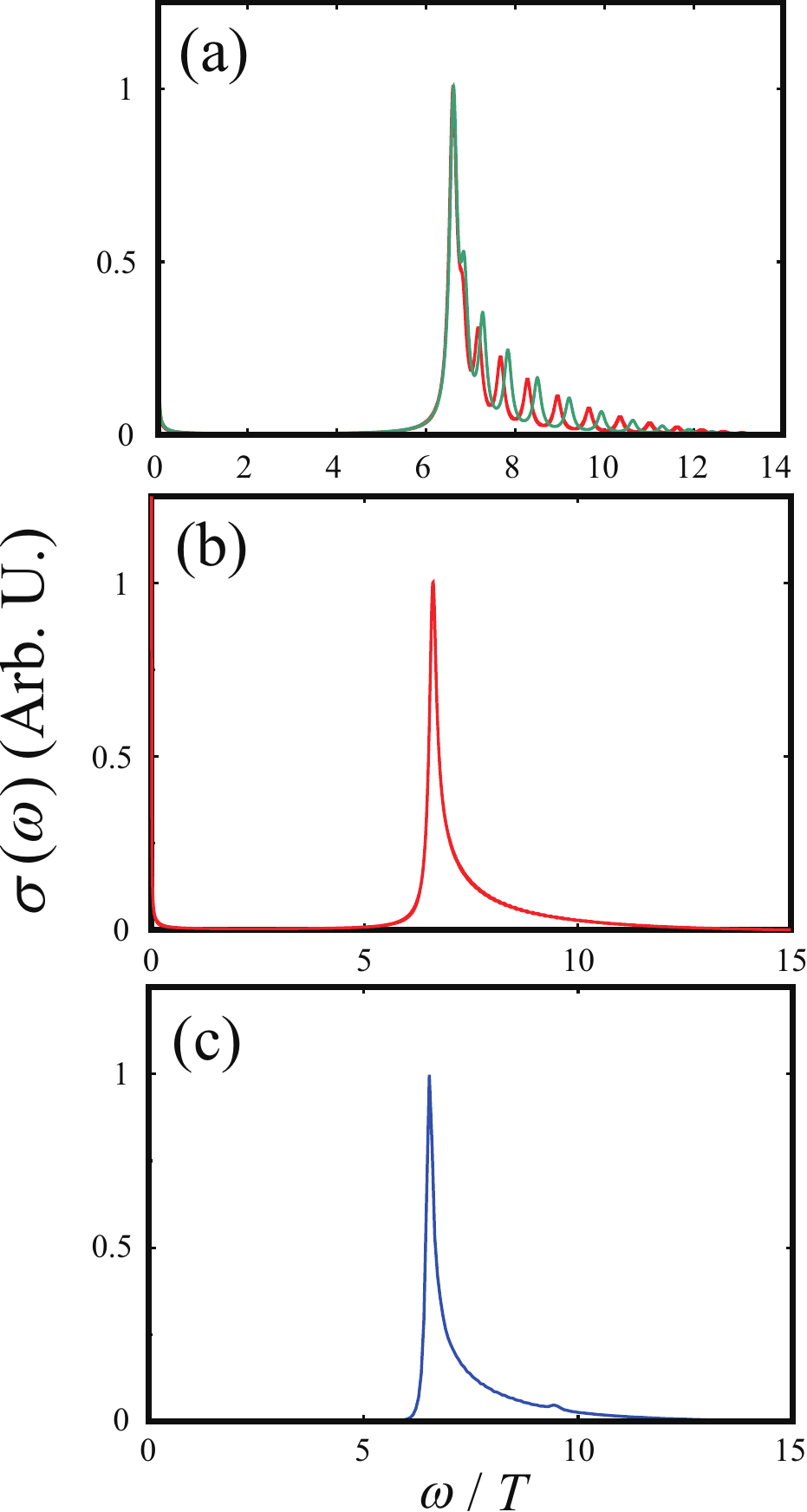}
	\caption{Optical conductivity spectra for $(U, V)/T=(10, 2.5)$. (a) Spectrum obtained with MBWFs (red line) and its comparison with that by 
	direct calculation with truncation (green line). Both are based on the charge model with $N$(or $N_{\rm ex}$)=40 and $\gamma=0.1T$.
	(b) Spectrum obtained with MBWFs for the charge model with $N_{\rm ex}=200$. (c) 
	Spectrum obtained using t-DMRG for the extended Hubbard model with $N=80$. 
	$\gamma$ is $0.10T$ and $0.14T$ for MBWF and t-DMRG, respectively.  Each intensity is normalized using its maximum.
	}
		\label{MBWF-8}
	\end{centering}
\end{figure}

\section{Summary and discussions} \label{sec:SD}
We have introduced an effective model, called the charge model, for the 1D Hubbard and extended Hubbard models, where spin-charge separation holds but charge fluctuations are not suppressed. First, using a finite ring, we found that the charge model reproduces the optical conductivity 
of the latter models satisfactorily in the intermediate and strong $U/T$ range. This shows that spin-charge separation holds quite nicely despite the significant charge fluctuation in the energy eigenstates that dominate the optical conductivity in the 1D Mott insulators of realistic correlation strength, and that this is the origin of their characteristic optical conductivity spectra.  Second, using the charge model, by the extrapolation of the Hamiltonian matrix to larger system sizes using MBWFs, we succeeded in calculating an almost-converged optical conductivity with respect to the system size. The optical conductivity spectra calculated using DDMRG and t-DMRG methods are reproduced well by the present method 
for sufficiently large systems in which finite-size effects are not significant.
The optical conductivity in the thermodynamic limit can be calculated effectively by the present method.

This enables us to compare the theoretical and experimental results directly even discussing a spectral shape. In contrast to DDMRG and t-DMRG methods, the present method yields the wave functions of the photoexcited states. The analysis of the calculated wave functions combined with the direct comparison with experiments provides a new viewpoint to understand the optical properties of strongly correlated electron systems. This problem is to be investigated in a forthcoming paper.

Only linear absorption spectra were considered, assuming a small vector potential, $A$, 
although quite interesting phenomena have been observed when strong excitations are present. For example, in experiments, the Mott gap was observed to be destroyed under intense photoexcitation.\cite{PIPT1} This annihilation of the Mott gap was shown to result from spin-charge coupling induced by intense photoexcitation.\cite{PIPTSCC} The spin-charge coupling in the intensely photoexcited states is the key to understand the origin of the photoinduced transition. It can be investigated by comparing the transient absorption spectra calculated from two models, namely, the original Hubbard and extended Hubbard, as well as the charge model, because the latter lacks the spin degrees of freedom.
This problem is also investigated in a forthcoming paper.

The present extrapolation method is applicable also to the Hubbard and extended Hubbard models
and to the strong excitation case. 
MBWFs in these models can have different spin structures in contrast to those in the charge model, 
and the analysis of the spin structures helps in understanding the spin-charge interaction in the 1D Mott insulators especially 
in the strong excitation case. Indeed, in this case, we need to consider multiple H-D pair basis states, and the number of important energy eigenstates in the starting cluster calculation becomes much larger as a result of spin-charge coupling. Furthermore, the construction of MBWFs is not straightforward and left for future study.

\section*{Acknowledgments}
This work was supported by JST CREST in Japan (Grant No. JPMJCR1661). K.I. was supported by the Grant-in-Aid for Scientific Research from JSPS in Japan (Grant No. JP17K05509). The computations were partially performed at Research Center for Computational Science, Okazaki, Japan,
and using supercomputers at RCNP and CMC of Osaka University, Osaka, Japan.

\appendix
\section{Transfer terms of the charge model} \label{app1}
In this appendix, we present the transfer terms ${\hat K}_n^{\rm (C)}(t)$ of the charge model.
The state $|\Phi_n^{\rm (C)}\rangle={\hat K}_n^{\rm (C)}(t)|\{p_1,p_2,\cdots,p_M\},\{q_1,q_2,\cdots,q_M\}\rangle$ for $1 \le n \le N-1$ is given by the following equations:

(i) Transfer of an H or D.

For $p_i=n$ and $l_k=n+1$,
\begin{eqnarray} \label{eq:CMH1}
|\Phi_n^{\rm (C)}\rangle=Te^{-iA(t)}|\{p_1,p_2,\cdots,p_{i-1},p_i+1,p_{i+1},\cdots,p_M\}\{q_1,q_2,\cdots,q_M\}\rangle,
\end{eqnarray}
for $l_k=n$ and $p_i=n+1$,
\begin{eqnarray} \label{eq:CMH2}
|\Phi_n^{\rm (C)}\rangle=-Te^{iA(t)}|\{p_1,p_2,\cdots,p_{i-1},p_i-1,p_{i+1},\cdots,p_M\}\{q_1,q_2,\cdots,q_M\}\rangle,
\end{eqnarray}
for $q_j=n$ and $l_k=n+1$,
\begin{eqnarray} \label{eq:CMH3}
|\Phi_n^{\rm (C)}\rangle=-Te^{iA(t)}|\{p_1,p_2,\cdots,p_M\}\{q_1,q_2,\cdots,q_{j-1},q_j+1,q_{j+1},\cdots,q_M\}\rangle,
\end{eqnarray}
for $l_k=n$ and $q_j=n+1$,
\begin{eqnarray} \label{eq:CMH4}
|\Phi_n^{\rm (C)}\rangle=-Te^{-iA(t)}|\{p_1,p_2,\cdots,p_M\}\{q_1,q_2,\cdots,q_{j-1},q_j-1,q_{j+1},\cdots,q_M\}\rangle.
\end{eqnarray}

(ii) Annihilation of an H-D pair.

For $p_i=n$, $q_j=n+1$, $l_k<p_i$, and $q_j<l_{k+1}$,
\begin{eqnarray} \label{eq:CMH6}
|\Phi_n^{\rm (C)}\rangle &=& -{\sqrt 2}Tc_{\rm S}(M)e^{-iA(t)}e^{-ik(\theta_M-\theta_{M-1})} \nonumber \\
&\times& |\{p_1,p_2,\cdots,p_{i-1},p_{i+1},\cdots,p_M\}\{q_1,q_2,\cdots,q_{j-1},q_{j+1},\cdots,q_M\}\rangle,
\end{eqnarray}
for $q_j=n$, $p_i=n+1$, $l_k<q_j$, and $p_i<l_{k+1}$,
\begin{eqnarray} \label{eq:CMH7}
|\Phi_n^{\rm (C)}\rangle &=& -{\sqrt 2}Tc_{\rm S}(M)e^{iA(t)}e^{-ik(\theta_M-\theta_{M-1})} \nonumber \\
&\times& |\{p_1,p_2,\cdots,p_{i-1},p_{i+1},\cdots,p_M\}\{q_1,q_2,\cdots,q_{j-1},q_{j+1},\cdots,q_M\}\rangle.
\end{eqnarray}

(iii) Creation of an H-D pair.

For $l_k=n$, $l_{k+1}=n+1$, $p_i<n$, $n+1<p_{i+1}$, $q_j<n$, and $n+1<q_{j+1}$,
\begin{eqnarray} \label{eq:CMH5}
|\Phi_n^{\rm (C)}\rangle &=&-{\sqrt 2}Tc_{\rm S}(M)e^{-i(k-1)(\theta_M-\theta_{M+1})} \nonumber \\
&\times& [e^{iA(t)}|\{p_1,p_2,\cdots,p_i,l_k,p_{i+1},\cdots,p_M\}\{q_1,q_2,\cdots,q_j,l_{k+1},q_{j+1},\cdots,q_M\}\rangle \nonumber \\
&+& e^{-iA(t)}|\{p_1,p_2,\cdots,p_i,l_{k+1},p_{i+1},\cdots,p_M\}\{q_1,q_2,\cdots,q_j,l_k,q_{j+1},\cdots,q_M\}\rangle].
\end{eqnarray}

The state $|\Phi_N^{\rm (C)}\rangle={\hat K}_N^{\rm (C)}(t)|\{p_1,p_2,\cdots,p_M\},\{q_1,q_2,\cdots,q_M\}\rangle$ is given by the following equations:

(i) Transfer of an H or D.

For $p_M=N$ and $l_1=1$,
\begin{eqnarray} \label{eq:CMHN1}
|\Phi_N^{\rm (C)}\rangle=-Te^{-iA(t)}e^{-i\theta_M}|\{1,p_1,\cdots,p_{M-1}\}\{q_1,q_2,\cdots,q_M\}\rangle,
\end{eqnarray}
for $l_{N-2M}=N$ and $p_1=1$,
\begin{eqnarray} \label{eq:CMHN2}
|\Phi_N^{\rm (C)}\rangle=-Te^{iA(t)}e^{i\theta_M}|\{p_2,\cdots,p_M,N\}\{q_1,q_2,\cdots,q_M\}\rangle,
\end{eqnarray}
for $q_M=N$ and $l_1=1$,
\begin{eqnarray} \label{eq:CMHN3}
|\Phi_N^{\rm (C)}\rangle=Te^{iA(t)}e^{-i\theta_M}|\{p_1,p_2,\cdots,p_M\}\{1,q_1,\cdots,q_{M-1}\}\rangle,
\end{eqnarray}
for $l_{N-2M}=N$ and $q_1=1$,
\begin{eqnarray} \label{eq:CMHN4}
|\Phi_N^{\rm (C)}\rangle=Te^{-iA(t)}e^{i\theta_M}|\{p_1,p_2,\cdots,p_M\}\{q_2,\cdots,q_M,N\}\rangle.
\end{eqnarray}

(ii) Annihilation of an H-D pair.

for $p_M=N$ and $q_1=1$,
\begin{eqnarray} \label{eq:CMHN6}
|\Phi_N^{\rm (C)}\rangle={\sqrt 2}Tc_{\rm S}(M)e^{-iA(t)}e^{-i\theta_{M-1}}|\{p_1,p_2,\cdots,p_{M-1}\}\{q_2,q_3,\cdots,q_M\}\rangle,
\end{eqnarray}
for $q_M=N$ and $p_1=1$,
\begin{eqnarray} \label{eq:CMHN7}
|\Phi_N^{\rm (C)}\rangle={\sqrt 2}Tc_{\rm S}(M)e^{iA(t)}e^{-i\theta_{M-1}}|\{p_2,p_3,\cdots,p_M\}\{q_1,q_2,\cdots,q_{M-1}\}\rangle.
\end{eqnarray}

(iii) Creation of an H-D pair.

For $l_{N-2M}=N$ and $l_1=1$,
\begin{eqnarray} \label{eq:CMHN5}
|\Phi_N^{\rm (C)}\rangle &=& {\sqrt 2}Tc_{\rm S}(M)e^{i\theta_M} \nonumber \\
&\times& [e^{iA(t)}|\{p_1,\cdots,p_M,N\}\{1,q_1,\cdots,q_M\}\rangle \nonumber \\
&+& e^{-iA(t)}|\{1,p_1,\cdots,p_M\}\{q_1,\cdots,q_M,N\}\rangle].
\end{eqnarray}

\section{Time-dependent DMRG} \label{app2}
We briefly explain the time-dependent density matrix renormalization group (t-DMRG), which is used for the benchmark calculation of our new model.
The dynamics of wave function $|\psi (t)\rangle$ of quantum systems is described by the time-dependent Schr\"{o}dinger equation, whose solution is given by
\begin{eqnarray}
    |\psi (t)\rangle = U(t,0)|\psi(0) \rangle ,
\end{eqnarray}
where $|\psi (0)\rangle$ is the wave function at initial time $t=0$.
Here, 
\begin{eqnarray}
    U(t,0)=T \exp \left[ -i\int _0 ^t ds H(s) \right]
\end{eqnarray}
is the time-evolution operator with the time-ordering operator $T$ and the time-dependent Hamiltonian $H(t)$.
For small time step $dt$, we can approximate 
\begin{eqnarray}
    U(t+dt,t)\simeq \exp [-idtH(t)].
\end{eqnarray}
To obtain $|\psi(t)\rangle$ accurately, we need to calculate $U(t+dt,t)$ as precise as possible.
One of the efficient approximations for $U(t+dt,t)$ is given by using the Suzuki-Trotter decomposition.\cite{tDMRG1}
However, this approach is basically restricted to the one-dimensional case.
Another approach is the use of the kernel polynomial method to approximate $U(t+dt,t)$ as follows.\cite{tDMRG2}
\begin{eqnarray}
    U(t+dt,t) = \sum_{l=0}^{\infty} (-i)^l (2l+1)j_l(dt)P_l(H(t)) \\
            \simeq \sum_{l=0}^{L} (-i)^l (2l+1)j_l(dt)P_l(H(t)),
\end{eqnarray}
where $j_l(s)$ is the spherical Bessel function of the first kind and $P_l(s)$ is the $l$-th Legendre polynomial.
They can be effectively obtained by the recurrence relations
\begin{eqnarray}
    j_{l+1}(x) = (2l+1)x^{-1}j_l(x) - j_{l-1}(x)
\end{eqnarray}
with $j_0(x)=x^{-1}\sin x$ and $j_1(x)=x^{-1}[-\cos x + x^{-1}\sin x]$ and
\begin{eqnarray}
    P_{l+1}(x) = \frac{2l+1}{l+1}xP_l(x) - \frac{l}{l+1}P_{l-1}(x)
\end{eqnarray}
with $P_0(x)=1$ and $P_1(x)=x$.
The calculation of the t-DMRG in the present study is performed by using the kernel polynomial method with the truncation number $L$, practically for $L\approx 10$, which gives a sufficiently converging result.
Furthermore, we use two target states $|\psi(t)\rangle$ and $|\psi(t+dt)\rangle$ in the t-DMRG procedure to effectively construct a basis that can express wave functions in time-dependent Hilbert space.
With the two-target t-DMRG procedure, we can calculate time-dependent physical quantities with high accuracy even when the Hamiltonian varies rapidly with time.

Calculating the time evolution of the current $J(t)$ induced by probe pulse by using t-DMRG, we obtained the optical conductivity $\sigma (\omega) =\frac{\tilde{J}(\omega)}{i(\omega + i\gamma)N\tilde{A}(\omega)}$, where $N$ is the system size, $\gamma$ is a broadening factor, $\tilde{J}(\omega)$ is the Fourier transform of $J(t)$, and $\tilde{A}(\omega)$ is the Fourier transform of the vector potential of the probe pulse $A(t)=A_0e^{-(t-t_0)^2/2t_d^2}\cos [\Omega (t-t_0)]$.\cite{tDMRG3,tDMRG4}
Here, the parameters of the probe pulse were $A_0=0.001$, $t_dT=0.02$, $\Omega/T=10$, and $t_0T=1$.
We employed open boundary conditions and kept 1000 density-matrix eigenstates.

\section{Matrix elements and their extrapolation in the case of finite $V$}\label{app3}
\begin{figure}[tb]
	\begin{centering}
	\includegraphics[width=0.6\textwidth]{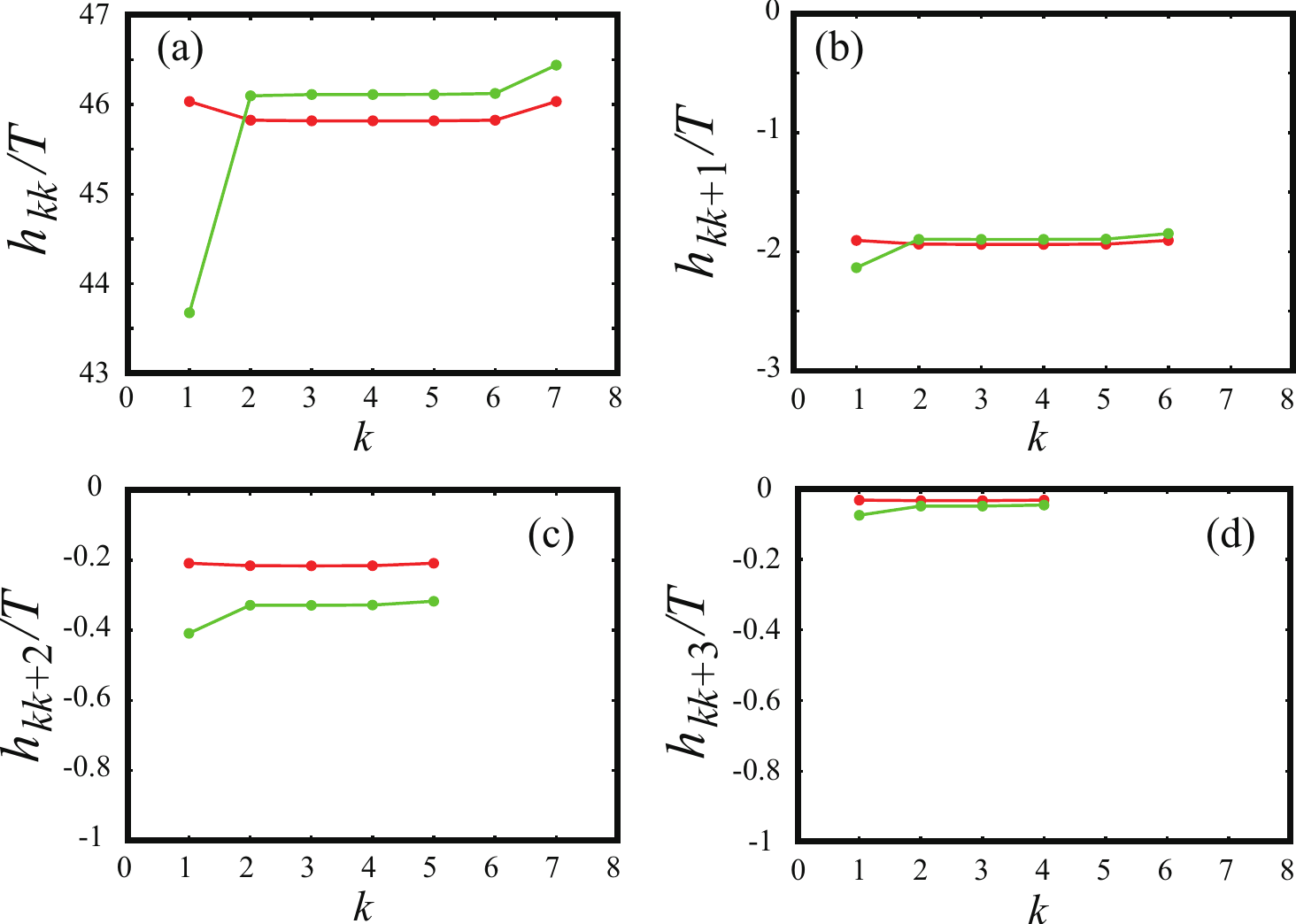}
	\caption{Matrix elements of $H^{\rm (C)}$ for $(U, V)/T=(10, 2.5)$ (green lines) and for $(U, V)/T=(10, 0)$ (red lines).}
		\label{MBWF-7}
	\end{centering}
\end{figure}

In, Fig.~\ref{MBWF-7}, we summarize the calculated matrix elements for $(U, V)/T=(10, 2.5)$.
For comparison, those for $(U, V)/T=(10, 0)$ are also shown.
Among the matrix elements, the most transparent effect 
due to the $V$ term appears as a sudden decrease at $k=1$ for the diagonal elements [Fig.~\ref{MBWF-7}(a)], 
which is naturally interpreted as an exciton effect.  We also notice significant deviations from the $V=0$ case
in the off-diagonal terms, particularly, in those of $h_{kk+2}$.  

In the extrapolation to distances $r_{\rm HD}\ge 8$, all such deviations are considered.  As actual procedures, we first replace
$h_{kk'}$ with an augmented matrix $(h_{\rm eff})_{kk'}$ defined as

\begin{eqnarray}
(h_{\rm eff})_{11} &=& h_{11}, \nonumber \\ 
(h_{\rm eff})_{kk} &=& m_0 \;\; (2 \le k \le I) \;,
\end{eqnarray}
\begin{eqnarray}
(h_{\rm eff})_{12} &=& h_{12}, \nonumber \\ 
(h_{\rm eff})_{kk+1} &=& m_1 \;\; (2 \le k\le I-1) \;,
\end{eqnarray}
\begin{eqnarray}
(h_{\rm eff})_{13} &=& h_{13}, \nonumber \\ 
(h_{\rm eff})_{kk+2} &=& m_2 \;\; (2 \le k\le I-2) \;,
\end{eqnarray}
and
\begin{eqnarray}
(h_{\rm eff})_{14} &=& h_{14}, \nonumber \\ 
(h_{\rm eff})_{kk+3} &=& m_3 \;\; (2 \le k\le I-3) \; ,
\end{eqnarray}
where the constants are averages defined as
\begin{eqnarray}
m_0 &=& {1\over5} \sum_{k=2}^6 h_{kk} \,  \nonumber \\
m_1 &=& {1\over5} \sum_{k=2}^6 h_{kk+1} \,  \nonumber  \\
m_2 &=& {1\over4} \sum_{k=2}^5 h_{kk+2} \,  \nonumber  \\
m_3 &=& {1\over3} \sum_{k=2}^4 h_{kk+3} \; .
\end{eqnarray}
Note that the undefined elements follow their symmetric counterparts if the latter are defined and zero otherwise.
Next, we redefine the effective model as $\tilde{h}_{\rm eff}\equiv  h_{\rm eff}-E_g^V$.  This is almost the same step
as that taken in the absence of $V$, although the change in the ground state energy arising from the $V$ term is considered
by replacing $E_g$ with $E_g^V \equiv E_g + \langle g| \hat{V}_V |g \rangle$, where $\hat{V}_V$ is the $V$ term.
Using this enlarged model, we calculated the optical conductivity spectrum based on the expression in Eq.~(\ref{eq:MBWF}).
Regarding the matrix elements of the current operator, we assume those elements defined in the absence of $V$, that is, 
those in Fig.~\ref{MBWF-3}(d).


\begin{thebibliography}{99}
\bibitem{TLL1} S. Tomonaga, Prog. Theor. Phys. {\bf 5}, 544 (1950).
\bibitem{TLL2} J. M. Luttinger, J. Math. Phys. {\bf 4}, 1154 (1963).
\bibitem{TLL3} D. C. Mattis and E. H. Lieb, J. Math. Phys. {\bf 6}, 304 (1965).
\bibitem{TLL4} J. S{\'o}lyom, Adv. Phys. {\bf 28}, 201 (1979).
\bibitem{TLL5} F. D. M. Haldane, J. Phys. C: Solid State Phys. {\bf 14}, 2585 (1981).
\bibitem{TLL6} H. J. Schulz, Int. J. Mod. Phys. B{\bf 5}, 57 (1991).
\bibitem{TLL7} J. Voit, Rep. Prog. Phys. {\bf 58}, 977 (1994).
\bibitem{SCSWF} M. Ogata and H. Shiba, Phys. Rev. B {\bf  41}, 2326 (1990).
\bibitem{SCSPL1} A. Parola and S. Sorella, Phys. Rev. Lett. {\bf 64}, 1831 (1990). 
\bibitem{SCSPL2} M. Ogata, T. Sugiyama, and H. Shiba, Phys. Rev. B {\bf  43}, 8401 (1991).
\bibitem{ARPES} B. J. Kim, H. Koh, E. Rotenberg, S.-J. Oh, H. Eisaki, N. Motoyama, S. Uchida, T, Tohyama, S. Maekawa, Z.-X. Shen, and C. Kim, Nature Physics {\bf 2}, 397 (2006).
\bibitem{NLOS1} H. Kishida, H. Matsuzaki, H. Okamoto, T. Manabe, M. Yamashita, Y. Taguchi, and Y. Tokura, Nature (London) {\bf 405}, 929 (2000).
\bibitem{NLOS2} Y. Mizuno, K. Tsutsui, T. Tohyama, and S. Maekawa, Phys. Rev. B {\bf  62}, R4769 (2000).
\bibitem{NLOS3} H. Kishida, M. Ono, K. Miura, H. Okamoto, M. Izumi, T. Manako, M. Kawasaki, Y. Taguchi, Y. Tokura, T. Tohyama, K. Tsutsui, and S. Maekawa, Phys. Rev. Lett. {\bf 87}, 177401 (2001).
\bibitem{NLOS4} M. Ono, K. Miura, A. Maeda, H. Matsuzaki, H. Kishida, Y. Taguchi, Y. Tokura, M. Yamashita, and H. Okamoto, Phys. Rev. B {\bf  70}, 085101 (2004).
\bibitem{PIPT1} S. Iwai, M. Ono, A. Maeda, H. Matsuzaki, H. Kishida, H. Okamoto, and Y. Tokura, Phys. Rev. Lett. {\bf 91}, 057401 (2003).
\bibitem{PIPT2} H. Okamoto, H. Matsuzaki, T. Wakabayashi, Y. Takahashi, and T. Hasegawa, Phys. Rev. Lett. {\bf 98}, 037401 (2007).
\bibitem{SCSEX1} F. Woynarovich, J. Phys. C: Solid State Phys. {\bf 15}, 85 (1982).
\bibitem{SCSEX2} A. Parola and S. Sorella, Phys. Rev. B {\bf 45}, 13156 (1992).
\bibitem{SCSOC1} H. Eskes, A. M. Ole\'{s}, M. B. J. Meinders, and W. Stephan, Phys. Rev. B {\bf 50}, 17980 (1994).
\bibitem{SCSOC2} K. Penc, K. Hallberg, F. Mila, and H. Shiba, Phys. Rev. B {\bf 55}, 15475 (1997).
\bibitem{stephan} W. Stephan and K. Penc, Phys. Rev. B {\bf 54}, R17269 (1996).
\bibitem{FinU1} S. Sorella and A. Parola, Phys. Rev. Lett. {\bf 76}, 4604 (1996).
\bibitem{FinU2} S. Sorella and A. Parola, Phys. Rev. B {\bf 57}, 6444 (1998).
\bibitem{FinU3} V. Lante and A. Parola, Phys. Rev. B {\bf 80}, 195113 (2009).
\bibitem{ED} E. Dagotto, Rev. Mod. Phys. {\bf 66}, 763 (1994).
\bibitem{DMRG} S. R. White, Phys. Rev. Lett. {\bf 69}, 2863 (1992).
\bibitem{DDMRG} E. Jeckelmann, F. Gebhard, F. H. L. Essler, Phys, Rev. Lett. {\bf 85}, 3910 (2000).
\bibitem{DDMRG2} F. H. L. Essler, F. Gebhard, and E. Jeckelmann, Phys. Rev. B {\bf 64}, 125119 (2001). 
\bibitem{DDMRG3} E. Jeckelmann, Phys. Rev. B {\bf 66}, 045114 (2002).
\bibitem{DDMRG4} E. Jeckelmann, Phys. Rev. B {\bf 67}, 075106 (2003).
\bibitem{DDMRG5} H. Benthien, F. Gebhard, and E. Jeckelmann, Phys. Rev. Lett. {\bf 92}, 256401 (2004).
\bibitem{DDMRG6} H. Benthien, and E. Jeckelmann, Phys. Rev. B {\bf 75}, 205128 (2007).
\bibitem{DDMRG7}S. S. Kancharla and C. J. Bolech, Phys. Rev. B {\bf 64}, 085119 (2001).
\bibitem{DDMRG8}A. C. Tiegel, T. Veness, P. E. Dargel, A. Honecker, T. Pruschke, I. P. McCulloch, and F. H. L. Essler, Phys. Rev. B {\bf 93}, 
125108 (2016).
\bibitem{OLHG} J. C. Talstra, S. P. Strong, and P. W. Anderson, Phys. Rev. Lett. {\bf 74}, 5256 (1995).
\bibitem{QMC1} H. Matsueda, N. Bulut, T. Tohyama,and S. Maekawa, Phys. Rev B {\bf 72}, 075136 (2005).
\bibitem{QMC2} A. S. Mishchenko and N. Nagaosa, Phys. Rev. Lett. {\bf 93}, 036402 (2004). 
\bibitem{PIPTSCC} A. Takahashi, H. Itoh, and M. Aihara, Phys. Rev. B {\bf 77}, 205105 (2008).
\bibitem{tDMRG1} S. R. White, and A. E. Feiguin, Phys. Rev. Lett. {\bf 93}, 076401 (2004).
\bibitem{tDMRG2} S. Sota, and M. Itoh, J. Phys. Soc. Jpn. {\bf 76}, 054004 (2007).
\bibitem{tDMRG3} H. Lu, C. Shao, J. Bon\v{c}a, D. Manske, and T. Tohyama, Phys. Rev. B {\bf 91}, 245117 (2015).
\bibitem{tDMRG4} C. Shao, T. Tohyama, H.-G. Luo, and H. Lu, Phys. Rev. B {\bf 93}, 195144 (2016).
\end{thebibliography}
\end{document}